\documentclass[%
reprint,
superscriptaddress,
nofootinbib,
notitlepage,
amsmath,
amssymb,
sortandcompress,
prd
]{revtex4-1}
\RequirePackage{lineno}

\usepackage[dvipsnames]{xcolor}
\usepackage{graphicx}
\graphicspath{{Fig/}}
\usepackage{dcolumn}
\usepackage{bm}
\usepackage{hyperref}
\usepackage{epsfig,amsmath,amssymb,verbatim,mathrsfs,array,layout,textcomp,amssymb,latexsym,slashed,booktabs,color,mathtools,tikz}
\usepackage{siunitx}
\usepackage{multirow}

\newcommand{\cevns}{CE$\nu$NS}

\def\mohm{m$\Omega$}
\def\tes{\mathrm{TES}}

\begin{document}

\title{Design and Characterization of a Phonon-Mediated Cryogenic Particle Detector with an eV-Scale Threshold and 100~keV-Scale Dynamic Range}

\affiliation{Department of Physics \& Astronomy, Northwestern University, Evanston, IL 60208-3112, USA}
\affiliation{Department of Physics, University of Florida, Gainesville, FL, USA}
\affiliation{Division of Physics, Mathematics, \& Astronomy, California Institute of Technology, Pasadena, CA 91125, USA}
\affiliation{Department of Physics, University of California, Berkeley, CA 94720, USA}
\affiliation{Department of Physics, University of Toronto, Toronto, ON M5S 1A7, Canada}
\affiliation{Fermi National Accelerator Laboratory, Batavia, IL 60510, USA} 
\affiliation{Kavli Institute for Cosmological Physics, University of Chicago, Chicago, IL 60637, USA}
\affiliation{School of Physics \& Astronomy, University of Minnesota, Minneapolis, MN 55455, USA}
\affiliation{Institut f\"ur Experimentalphysik, Universit\"at Hamburg, 22761 Hamburg, Germany}

\affiliation{Pacific Northwest National Laboratory, Richland, WA 99352, US}
\affiliation{Department of Physics and Astronomy, and the Mitchell Institute for Fundamental Physics and Astronomy, Texas A\&M University, College Station, TX 77843, USA}
\affiliation{Department of Physics, Stanford University, Stanford, CA 94305 USA}

\author{R.~Ren} \affiliation{Department of Physics \& Astronomy, Northwestern University, Evanston, IL 60208-3112, USA}
\author{C.~Bathurst} \affiliation{Department of Physics, University of Florida, Gainesville, FL, USA}
\author{Y.Y.~Chang} \affiliation{Division of Physics, Mathematics, \& Astronomy, California Institute of Technology, Pasadena, CA 91125, USA}
\author{R.~Chen} \affiliation{Department of Physics \& Astronomy, Northwestern University, Evanston, IL 60208-3112, USA}
\author{C.W.~Fink} \affiliation{Department of Physics, University of California, Berkeley, CA 94720, USA}
\author{Z.~Hong} \affiliation{Department of Physics \& Astronomy, Northwestern University, Evanston, IL 60208-3112, USA} \affiliation{Department of Physics, University of Toronto, Toronto, ON M5S 1A7, Canada} 
\author{N.A.~Kurinsky}\email{kurinsky@fnal.gov} \affiliation{Fermi National Accelerator Laboratory, Batavia, IL 60510, USA} \affiliation{Kavli Institute for Cosmological Physics, University of Chicago, Chicago, IL 60637, USA}
\author{N.~Mast} \affiliation{School of Physics \& Astronomy, University of Minnesota, Minneapolis, MN 55455, USA}
\author{N.~Mishra} \affiliation{Fermi National Accelerator Laboratory, Batavia, IL 60510, USA}
\author{V.~Novati}\email{valentina.novati@northwestern.edu}  \affiliation{Department of Physics \& Astronomy, Northwestern University, Evanston, IL 60208-3112, USA}
\author{G. Spahn} \affiliation{Fermi National Accelerator Laboratory, Batavia, IL 60510, USA}
\author{H.~Meyer zu Theenhausen} \affiliation{Institut f\"ur Experimentalphysik, Universit\"at Hamburg, 22761 Hamburg, Germany}
\author{S.L.~Watkins} \affiliation{Department of Physics, University of California, Berkeley, CA 94720, USA}
\author{Z.~Williams} \affiliation{School of Physics \& Astronomy, University of Minnesota, Minneapolis, MN 55455, USA}
\author{M.J.~Wilson} \affiliation{Institut f\"ur Experimentalphysik, Universit\"at Hamburg, 22761 Hamburg, Germany} \affiliation{Department of Physics, University of Toronto, Toronto, ON M5S 1A7, Canada} 
\author{A.~Zaytsev} \affiliation{Institut f\"ur Experimentalphysik, Universit\"at Hamburg, 22761 Hamburg, Germany}

\author{D.~Bauer} \affiliation{Fermi National Accelerator Laboratory, Batavia, IL 60510, USA}
\author{R.~Bunker} \affiliation{Pacific Northwest National Laboratory, Richland, WA 99352, US}
\author{E.~Figueroa-Feliciano} \affiliation{Department of Physics \& Astronomy, Northwestern University, Evanston, IL 60208-3112, USA}
\author{M.~Hollister} \affiliation{Fermi National Accelerator Laboratory, Batavia, IL 60510, USA}
\author{L. Hsu} \affiliation{Fermi National Accelerator Laboratory, Batavia, IL 60510, USA}
\author{P.~Lukens} \affiliation{Fermi National Accelerator Laboratory, Batavia, IL 60510, USA}
\author{R.~Mahapatra} \affiliation{Department of Physics and Astronomy, and the Mitchell Institute for Fundamental Physics and Astronomy, Texas A\&M University, College Station, TX 77843, USA}
\author{N.~Mirabolfathi} \affiliation{Department of Physics and Astronomy, and the Mitchell Institute for Fundamental Physics and Astronomy, Texas A\&M University, College Station, TX 77843, USA}
\author{B.~Nebolsky} \affiliation{Department of Physics \& Astronomy, Northwestern University, Evanston, IL 60208-3112, USA}
\author{M.~Platt} \affiliation{Department of Physics and Astronomy, and the Mitchell Institute for Fundamental Physics and Astronomy, Texas A\&M University, College Station, TX 77843, USA}
\author{F.~Ponce} \affiliation{Department of Physics, Stanford University, Stanford, CA 94305 USA}
\author{M.~Pyle} \affiliation{Department of Physics, University of California, Berkeley, CA 94720, USA}
\author{T.~Reynolds} \affiliation{Department of Physics, University of Florida, Gainesville, FL, USA}
\author{T.~Saab} \affiliation{Department of Physics, University of Florida, Gainesville, FL, USA}

\date{\today}
\begin{abstract}
    We present the design and characterization of a cryogenic phonon-sensitive 1-gram Si detector exploiting the Neganov-Trofimov-Luke effect to detect single-charge excitations. This device achieved 2.65(2)~eV phonon energy resolution when operated without a voltage bias across the crystal and a corresponding charge resolution of 0.03 electron-hole pairs at 100~V bias. With a continuous-readout data acquisition system and an offline optimum-filter trigger, we obtain a 9.2~eV threshold with a trigger rate of the order of 20~Hz. The detector's energy scale is calibrated up to 120~keV using an energy estimator based on the pulse area. The high performance of this device allows its application to different fields where excellent energy resolution, low threshold, and large dynamic range are required, including dark matter searches, precision measurements of coherent neutrino-nucleus scattering, and ionization yield measurements. 
    
\end{abstract}

\maketitle

\section{Introduction}
Driven by the needs of both rare-event searches and coherent elastic neutrino-nucleus scattering (\cevns) experiments, substantial effort has been spent in refining the design of cryogenic calorimeters towards eV-scale energy thresholds. These efforts include cryogenic CCDs demonstrated by SENSEI~\cite{Tiffenberg:2017,Crisler:2018,Abramoff:2019,Barak:2020} and DAMIC~\cite{Aguilar-Arevalo:2019}, calorimetric detectors from CRESST~\cite{Abdelhameed:2019}, $\nu$-CLEUS~\cite{Strauss:2017,Angloher:2017}, EDELWEISS~\cite{Maisonobe:2018,Arnaud:2020,Armengaud:2019}, and SuperCDMS~\cite{Romani:2018,Agnese:2018,Amaral:2020,Hong:2020,Fink:2020,Fink:2020b,Alkhatib:2020}. Of these technologies, the detector presented in this paper is the first capable of operating with a low threshold at 0~V, allowing us to measure recoil energy, while also being able to measure quantized charges under application of a voltage bias.

Since the first demonstrations of single-charge sensitive cryogenic Si detectors~\cite{Romani:2018,Hong:2020}, we have systematically studied the detector design through dedicated device characterization, to understand how to improve the energy resolution and lower the energy threshold for rare-event searches. Motivated by an interest to measure the ionization yield of nuclear recoils, we have also explored ways to increase the dynamic range of these detectors to allow them to probe eV- to keV-scale energies. This paper presents the best resolution yet achieved for a gram-scale phonon-mediated calorimeter (2.65(2)~eV), the highest energy collection efficiency ($\gtrsim$29\%), and the highest calibrated dynamic range of a single-charge sensitive device (up to 120~keV). The measured performance has been achieved in multiple cryogenic systems and matches our model prediction from the design of the device well.

The organization of this paper is as follows. In Section~\ref{sec:det}, we briefly review the components of the energy resolution model for a generic athermal phonon detector. In Section~\ref{sec:design}, we apply this resolution model to the detector geometry discussed in this paper and discuss a new detector response model used to optimize the dynamic range (described in detail in Appendix~\ref{app:DR}). In Section~\ref{sec:expsetup}, we discuss the experimental setups in which this device was tested and in Section~\ref{sec:char} we combine the results of those tests to compare the performance to the detector response model. We present the event reconstruction algorithms employed in Section~\ref{sec:recon} and discuss their performance for events near threshold and at high energy in Section~\ref{sec:perf}. Finally, we summarize the main findings of this paper in Section~\ref{sec:conclusion}.

\section{QET Phonon Detectors}\label{sec:det}
\begin{figure}[!h]
    \centering
    \includegraphics[width=0.49\textwidth]{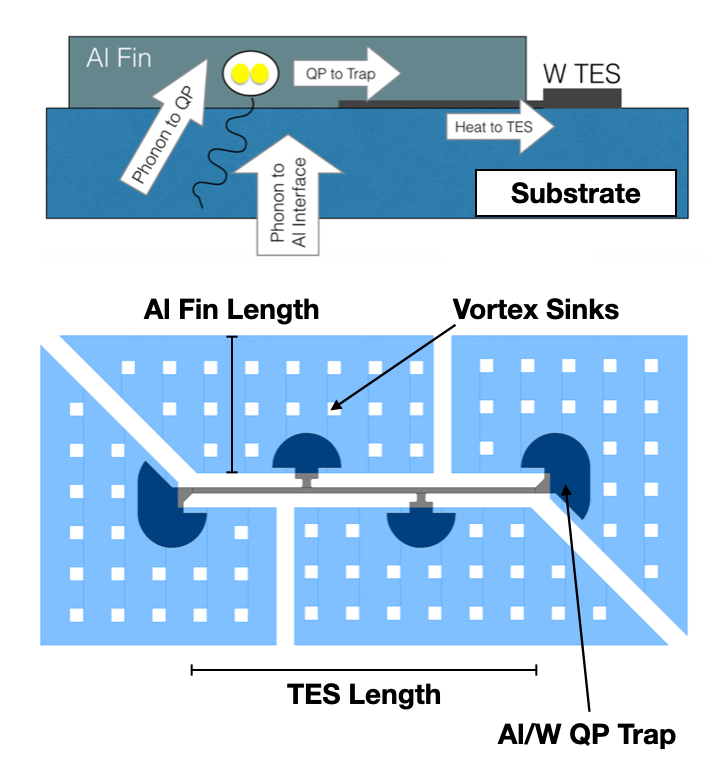}
    \caption{Overview of QET energy transport (top) and design geometry (bottom). 
    Athermal phonons generated by events in the substrate propagate with high efficiency to the Al/substrate interface, where they are either transmitted or reflected. The transmitted phonons break Cooper pairs in the Al creating free, athermal quasiparticles (QPs), which diffuse through the fin from the initial event. When these QPs encounter the lower-gap energy region of the Al/W quasiparticle trap, they convert most of the initial gap energy to phonons, heating the TES.}
    \label{fig:qet}
\end{figure}

Superconducting phonon calorimeters employ large target volumes coupled to smaller volume superconductors to channel energy into a small heat capacity that can be read out at high signal to noise ratio. Our design uses a parallel array of Quasiparticle-trap-assisted Electro-thermal-feedback Transition-edge sensors (QETs)~\cite{Irwin:1995} for each readout channel. As shown in Fig.~\ref{fig:qet}, a QET-based detector consists of three components: (1) a macroscopic substrate as the particle-sensing target (a Si crystal in this case), (2) a superconducting thin film as a phonon collector (the Al fins), and (3) a Transition-Edge Sensor (TES)~\cite{Irwin:2005}. 

The TES is made of W with a critical temperature tuned to $\sim$65~mK in the devices described in this paper. On a microscopic level, phonon energy in the target from particle interactions is converted to superconducting quasi-particles in the Al fin phonon collectors. The Al fins employ their small superconducting gap energy ($\sim$350~$\mathrm{\mu}$eV for Al) to separate athermal phonons from the residual thermal phonons at low temperature ($\sim1~\mathrm{\mu eV}$ at 10~mK), thus providing a relatively fast sensor response. The Al/W overlap region has a lower gap than the Al bulk, forming a quasiparticle trap which funnels quasiparticles into the much smaller TES volume as they shed energy via the emission of phonons. 

The TESs connected to these traps are operated in their superconducting transition with a voltage bias, producing an electro-thermal feedback effect~\cite{Irwin:2005}. They convert the phonon energy into a current change which can be sensed using cryogenic amplifiers. The parallel array of QET cells are spread out over the crystal surface, with the number of cells, coverage pattern, and individual QET design all affecting the performance of the device. 

As detailed in Refs.~\cite{Irwin:2005,Pyle:2012,Kurinsky:2018}, the intrinsic energy resolution of a TES microcalorimeter can be written in terms of the detector bandwidth (expressed as the time constant $\tau_{\mathrm{BW}}$), the efficiency of phonon energy collection $\epsilon$, the thermal conductance $G$ between the TES and the crystal (and associated power-law constant $n$), and the calorimeter operating temperature $T_0$ as
\begin{equation}\label{eq:sigma}
    \sigma_{ph} = \frac{1}{\epsilon}\sqrt{2Gk_bT_0^2\tau_{\mathrm{BW}}},
\end{equation}
where $k_b$ is the Boltzmann constant. For a TES with a narrow transition width, $T_0$ can be reasonably approximated by the critical temperature $T_c$ of the TES.

For these devices the $G$ is set by the volume of the TES and its electron-phonon coupling. The thermal power between the TES and the crystal is described by the equation
\begin{equation}
    P_0 = \Sigma \frac{v_\tes}{\zeta_\tes} T_c^n\left(1 - \left[\frac{T_b}{T_c}\right]^n\right)
\end{equation}
such that, when linearized around $T_c$, the thermal conductance is
\begin{equation}
    G \approx n\Sigma \frac{v_\tes}{\zeta_\tes}T_c^{n-1} \approx \left.\frac{nP_0}{T_c}\right|_{T_b << T_c},
\end{equation}
where $\Sigma$ is the electron-phonon coupling constant for a W TES, $\zeta_\tes$ is the fraction of the W volume contained in the TES length, $v_\tes$ is the total TES volume and $n$ is the thermal conductance power-law exponent for the power equation, nominally taken to be $n\sim 5$ for electron-phonon coupling~\cite{Giazotto:2006}. 
$T_b$ is the base temperature of the cryostat, also known as bath temperature. Because there is a strong thermal conductance between the crystal and the bath in our setup, we refer to the crystal temperature also as $T_b$ when the system is at equilibrium.
This allows us to substitute G in the resolution scaling, giving
\begin{equation}
    \sigma_{ph} \approx \frac{T_c^3}{\epsilon}\sqrt{2n\Sigma \frac{v_\tes}{\zeta_\tes}k_b\left(\tau_{ph}+\tau_{-}\right)},
    \label{eq:t3}
\end{equation}
where the bandwidth has been broken into phonon collection time $\tau_{ph}$ and effective TES response time $\tau_{-}$ (see e.g. Ref~\cite{Fink:2020})). This result for athermal phonon detectors shows that the energy resolution scales as $T_c^3$ when phonon dynamics limit the integration time and the TES is limited by its own thermal fluctuations. 

An additional consideration in detector design which becomes relevant for more general purpose TES detectors is dynamic range, and the related quantity, saturation energy. The resolution model described above applies strictly in the small-signal limit; away from this limit, the TES response becomes non-linear, and for large enough events, enough energy is supplied to drive the TES into the normal state, which is referred to as the saturation energy. For transition width $\Delta T_c$ and specific heat $c_{W}$ we find a saturation energy $E_{sat}$ of
\begin{equation}
    E_{sat} \approx \frac{1}{\epsilon}C(\Delta T_c) = \frac{1}{\epsilon}c_{W}\frac{v_{TES}}{\zeta_{TES}}T_c(\Delta T_c).
\end{equation}
We thus see that many of the design drivers that minimize resolution (e.g. reducing TES volume and bias power) also reduce saturation energy.
The total pulse integral is still a singular function of event energy above this point, but the saturation energy sets a rough scale where the TES goes from the linear to non-linear regime, and the resolution becomes energy dependent. The linear dynamic range is thus roughly the ratio of saturation energy to resolution, which scales as roughly
\begin{equation}
    DR \sim \frac{E_{sat}}{\sigma_{ph}} \propto \frac{\sqrt{v_{TES}}}{T_c^2\sqrt{\tau_{BW}}}(\Delta T_c)
\end{equation}
and we see that, for fixed $T_c$, smaller TES volume decreases overall dynamic range. The subject of this paper is largely how to balance the typical TES resolution model, summarized above, with the dynamic range model we present for the first time in this paper. We also compare the model predictions with the measured detector response. We make the model in this paragraph more precise by including the TES response model; a reader interested in that modeling can jump to Appendix~\ref{app:DR} before proceeding to the next section for more detail.

\section{Detector Optimization}\label{sec:design}
\label{sec:det_optimization}
The detector described in this paper (referred to as NF-C) was designed for ionization yield measurements in a neutron beam at the Triangle Universities Nuclear Laboratory (TUNL)~\cite{ZH_APS_talk}. This application required a device that could measure large energy depositions ($\sim100$~keV) while maintaining excellent baseline resolution.   
NF-C is a re-optimization of the detector mask from Ref.~\cite{Hong:2020} (referred to as QP.4), which attained the desired energy resolution (3~eV), but not the dynamic range.
We apply the modeling framework described in~\cite{Kurinsky:2018} to map out the response of detectors as we varied design parameters. 

\begin{figure*}[th]
    \centering
    \includegraphics[height=2.7in]{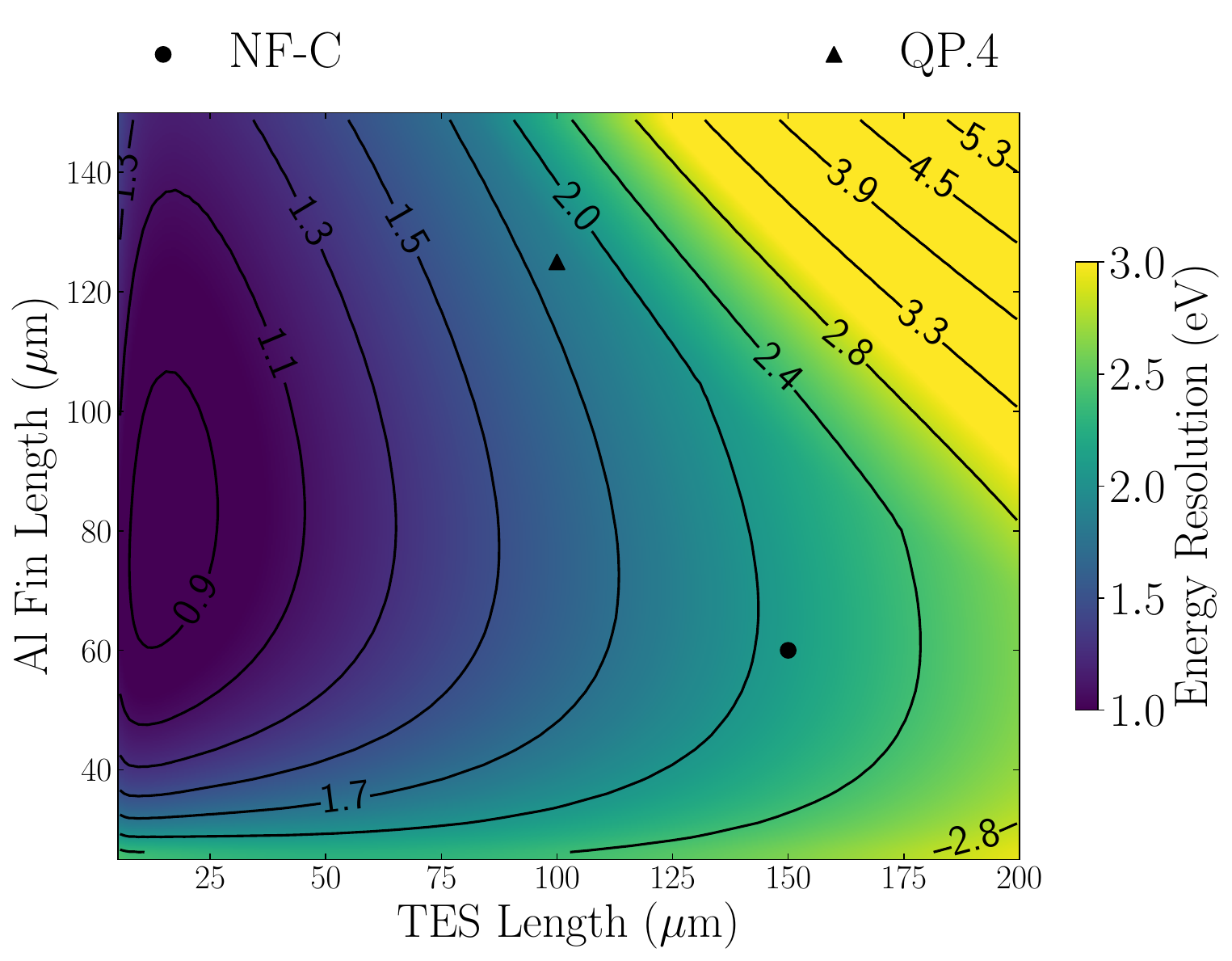}
    \includegraphics[height=2.7in]{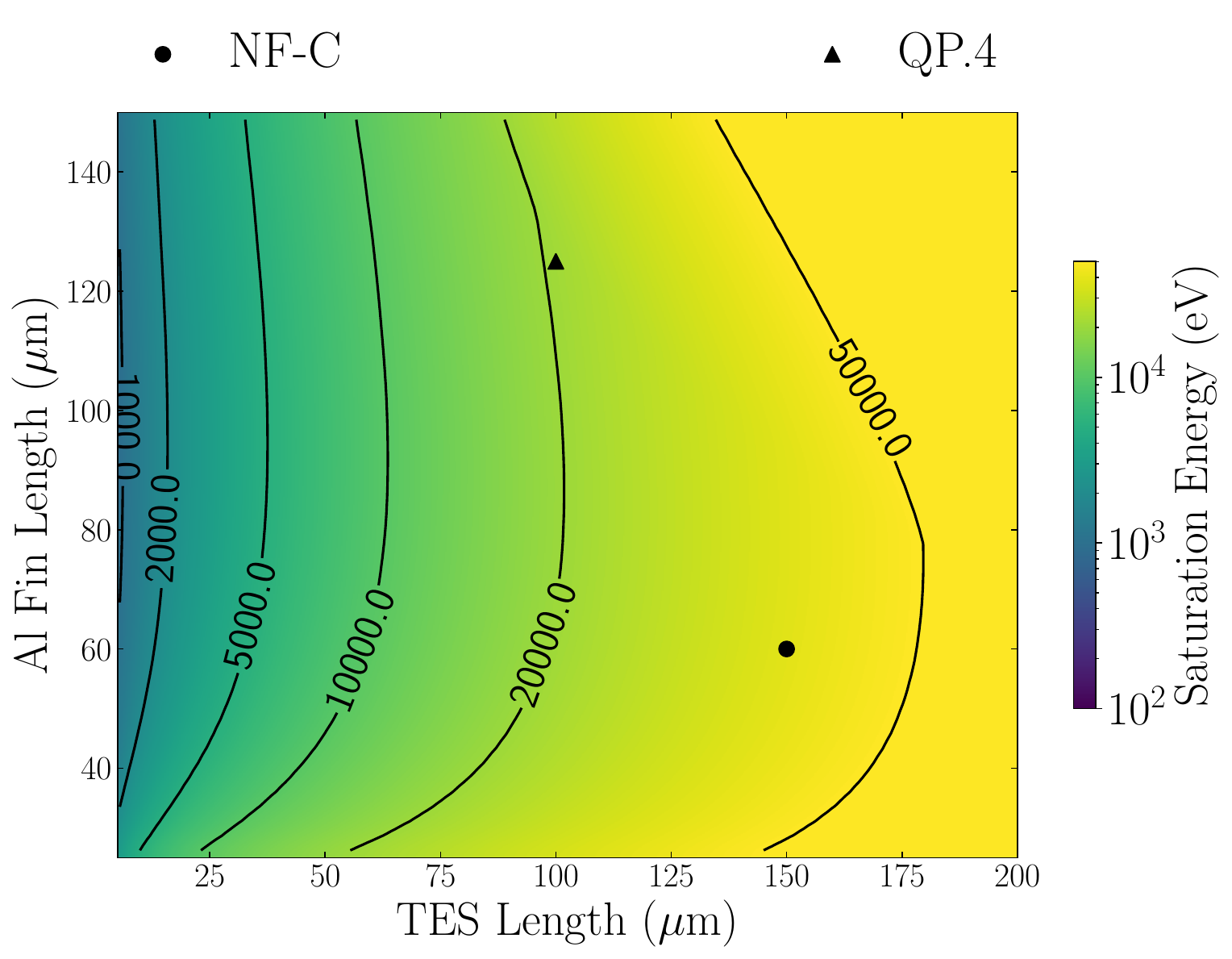}
    \includegraphics[height=2.7in]{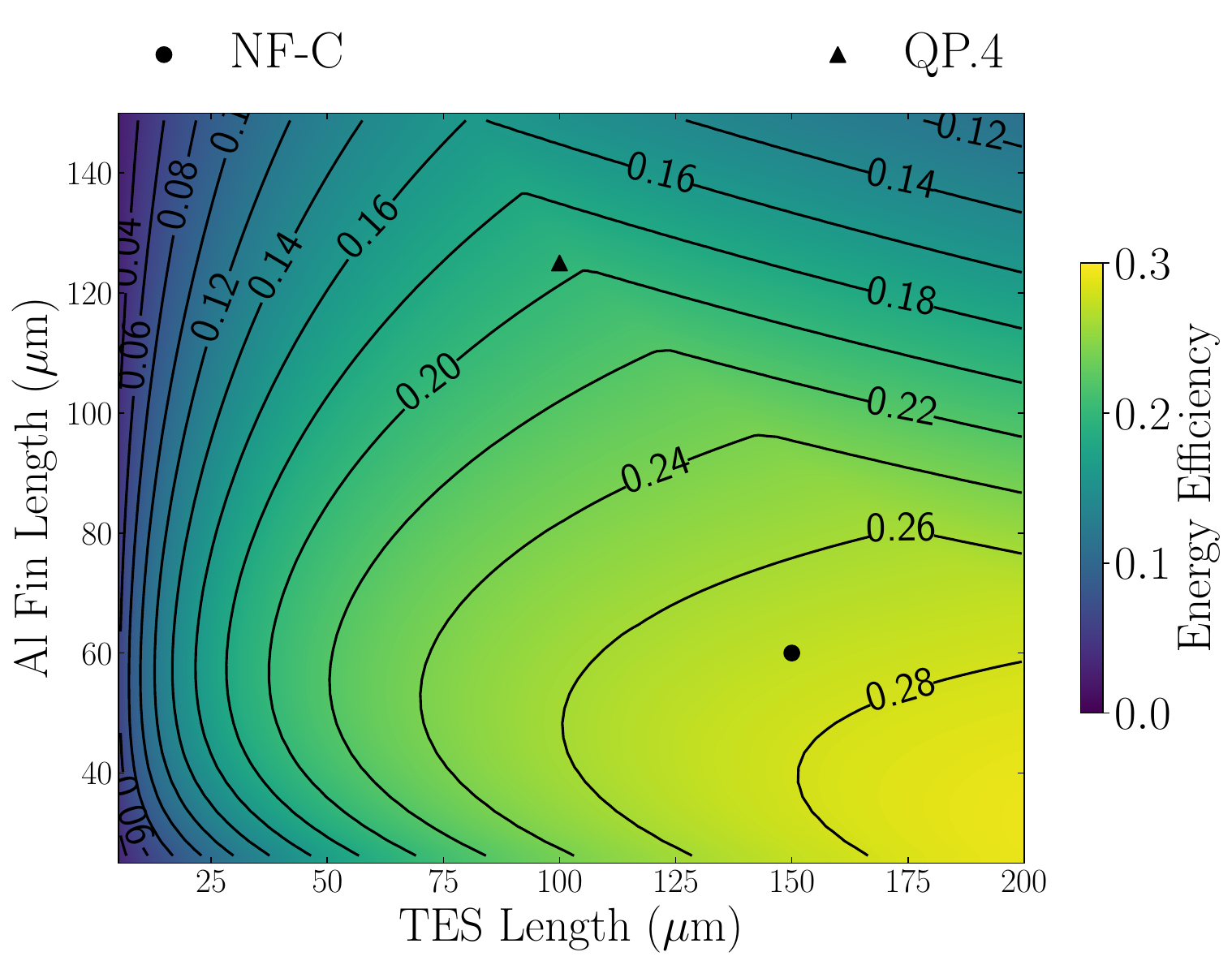}
    \includegraphics[height=2.7in]{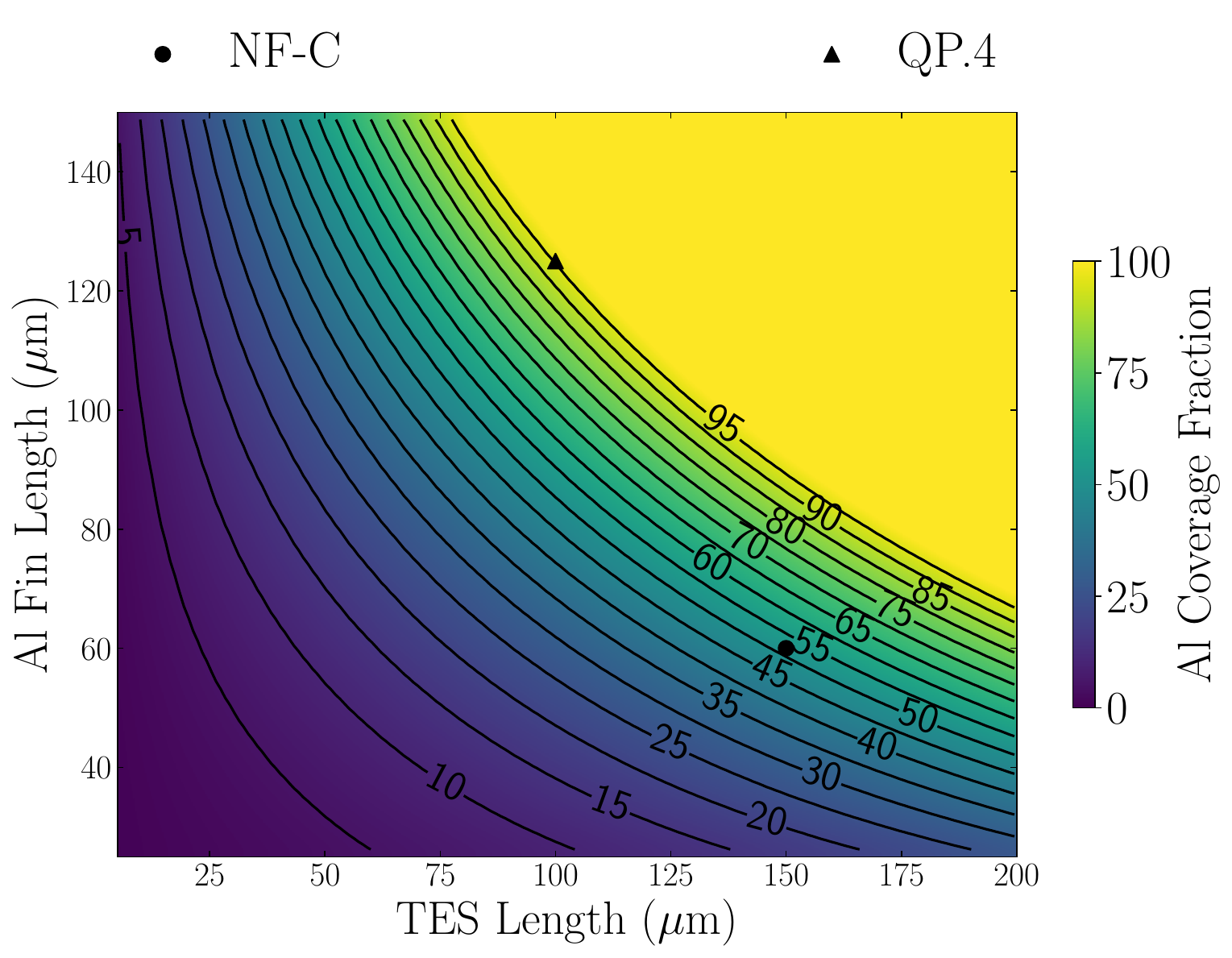}
    
    \caption{Top panel: Detector energy resolution (left) and saturation energy (right) as a function of the Al fin and TES lengths.
    Bottom panel: Energy efficiency (left) and Al coverage fraction (right) as a function of the Al fin and TES lengths. The model predictions for detectors QP.4 (triangle) and NF-C (circle) are also shown.}
    \label{fig:detmodel}
\end{figure*}

This detector response scan can be done independently of readout considerations by fixing the QET channel's overall normal resistance $R_n$. We also fix all TES properties (including $T_c$, width and thickness of the W) except for TES length ($l_\tes$) and Al fin length ($l_{fin}$) to those measured from QP.4~\cite{Hong:2020}. The number of QETs ($N_{\mathrm{QET}}$) in a channel is set to be a function of TES length such that $R_n$ is kept constant, with
\begin{equation}
    R_n = \frac{R_{\mathrm{TES,QET}}}{N_\mathrm{QET}}=\frac{\rho_\tes \cdot l_\tes}{w_\tes\cdot t_\tes \cdot N_\mathrm{QET}},
\end{equation}
where $\rho_\tes$ is the W resistivity (which is $T_c$ dependent), $w_\tes$ and $t_\tes$ are the width and thickness of the TES, respectively, and $R_\mathrm{TES,QET}$ is the normal resistance of each QET cell.
In this limit, the volume of TES per channel ($v_{\tes,ch} $) scales as 
\begin{equation}
    v_{\tes,ch} = N_\mathrm{QET} v_{\tes,\mathrm{QET}} = \frac{\rho_\tes}{R_n}l_\tes^2,
\end{equation}
where $v_{\tes,\mathrm{QET}}$ is the volume of the TES per QET cell. 

We parametrize the geometry of a QET cell in the 2-dimensional space of ($l_\tes$, $l_{fin}$). Because $l_\tes$ determines the number of QETs in a channel, the overall Al coverage fraction (a key parameter in the efficiency $\epsilon$) is also set by these two parameters. With these design rules we can parametrize the detector energy resolution, saturation energy, energy efficiency, and Al coverage fraction in the 2-dimensional space of ($l_\tes$, $l_{fin}$). The results of this modeling are shown in Fig.~\ref{fig:detmodel}, along with the design points for QP.4 and NF-C. The efficiency model from Ref.~\cite{Kurinsky:2018} is qualitatively described in Appendix~\ref{app:eff}.

As stated above, the NF-C design goal was to retain the QP.4 energy resolution while increasing the dynamic range, which is a function of the saturation energy. The dynamic range can be extended by increasing the volume of the TES (see Appendix~\ref{app:DR}). At the same time, from Eq.~\ref{eq:t3} we see that we can avoid degrading the energy resolution by simultaneously increasing the collection efficiency such that we keep the ratio $\sqrt{v_\tes}$/$\epsilon$ approximately constant. The chosen parameters for NF-C increase the efficiency projection from around 20\% to 27\% as the TES length increases from 100 to 150~$\mu$m, maintaining a relatively constant ratio of TES length to energy efficiency; as a result, the overall energy resolution is largely constant. The model predicts the dynamic range is increased by 50\% relative to the QP.4 detector. 

\subsection{Phonon-Assisted Charge Readout}

The detector ionization signal can be amplified by the Neganov-Trofimov-Luke (NTL) effect~\cite{Neganov:1985, Luke:1988}.
Initial electron-hole pairs are accelerated and drifted across the crystal in an electric field, resulting in an amplified phonon signal. The total phonon energy, $E_{ph}$, produced by the NTL effect is related to the initial energy deposition $E_r$ by:
\begin{equation}
    \label{eq:NTL}
    E_{ph}=E_{r}+n_{eh}\cdot e \cdot V_{\mathrm{NTL}} = E_{r}\left(1+\dfrac{e \cdot V_{\mathrm{NTL}}}{\varepsilon_\gamma(E_r)}\right),
\end{equation}
where $e$ is the elementary charge, $n_{eh}$ is the number of electron-hole pairs produced, $V_{\mathrm{NTL}}$ is the bias applied across the detector and  $\varepsilon_\gamma(E_r)$ is the average energy required to produce an electron-hole pair. While $\varepsilon_\gamma$ can be approximated by a constant 3.8~eV in silicon for high energy interactions, $\varepsilon_\gamma(E_r)$ is a function of the initial energy in the case of a few charge carriers~\cite{Koc:1957,Ramanathan:2020}. The signal can be amplified to the point that the detector is sensitive to a single electron-hole pair.

\begin{figure}[t]
    \centering
    \includegraphics[width=\columnwidth]{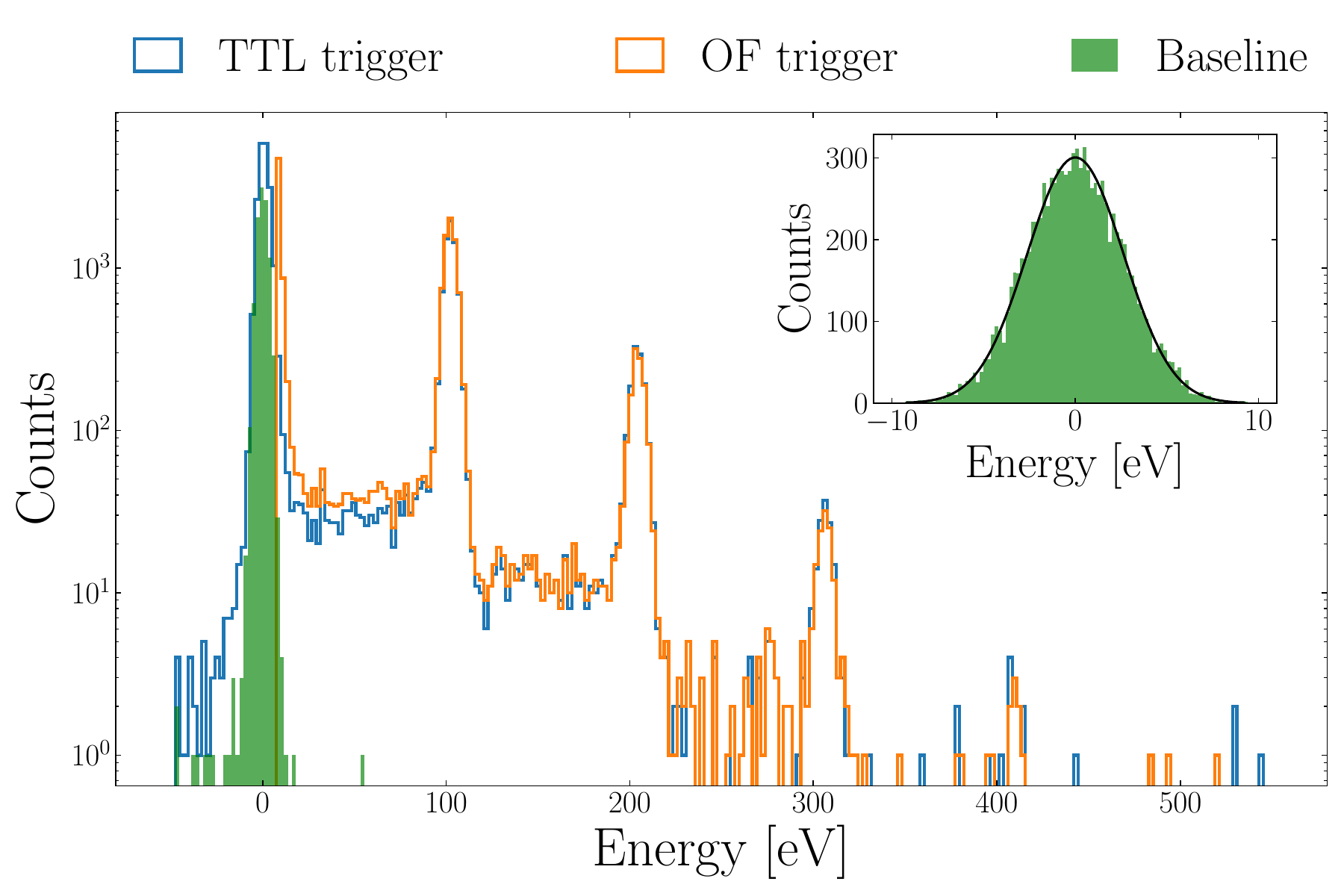}
    \caption{Laser distribution acquired with 100~V NTL bias. The data are triggered with the digital laser signal (TTL signal) and with the OF trigger. An energy resolution $\sigma =$ 3.25(4)~eV was measured at the first electron-hole pair peak. Inset: Zoom of the green histogram, which represents the random triggers used to estimate the baseline resolution. A baseline resolution $\sigma$ = 2.65(2)~eV was measured from random triggers. The Gaussian fit is shown with a black line}. The discrepancy between the baseline and peak resolution is due to additional variance from absorption of photons in the QETs~\cite{Hong:2020}.
    \label{fig:LowLambda}
\end{figure}

We can invert Eq.~\ref{eq:NTL} to obtain the charge resolution ($\sigma_q$) as a function of the phonon energy resolution ($\sigma_{ph}$) when the NTL amplification is significantly larger than the initial energy deposition:
\begin{equation}
    \sigma_{q}\approx \frac{\sigma_{ph}}{e\cdot V_{\mathrm{NTL}}}.
\end{equation}
Figure~\ref{fig:LowLambda} shows a calibration spectrum from laser data acquired at 100 V NTL bias that was used to evaluate the energy resolution. The energy includes both the initial energy deposited and the NTL amplification, therefore the first peak at 101.95 eV corresponds to an event generated by a single laser photon.
For a nominal voltage bias of 100~V and a phonon resolution of 3.25(4)~eV at the first electron-hole peak, we can therefore expect a charge resolution of $\sim$0.03 electron-hole pairs.

\section{Detector Operation and Experimental Setup}\label{sec:expsetup}

\begin{figure}[h]
    \centering
    \includegraphics[width=0.62\columnwidth]{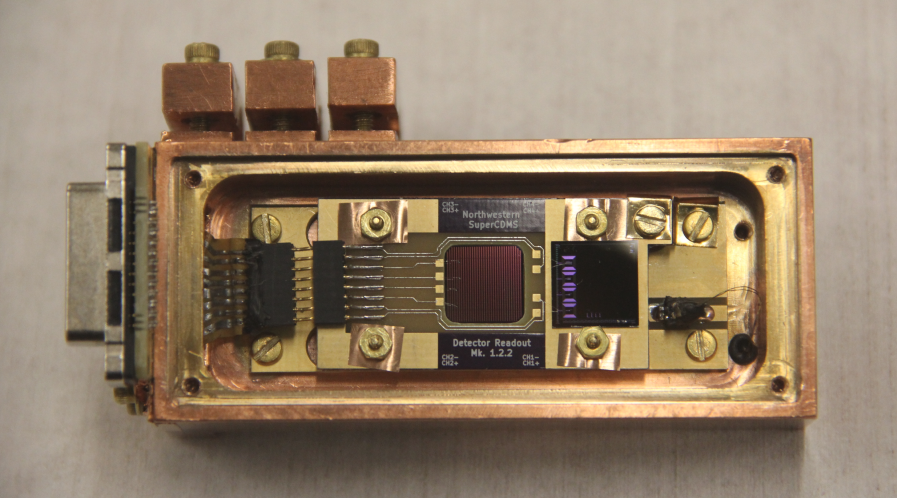}
    \includegraphics[width=0.36\columnwidth]{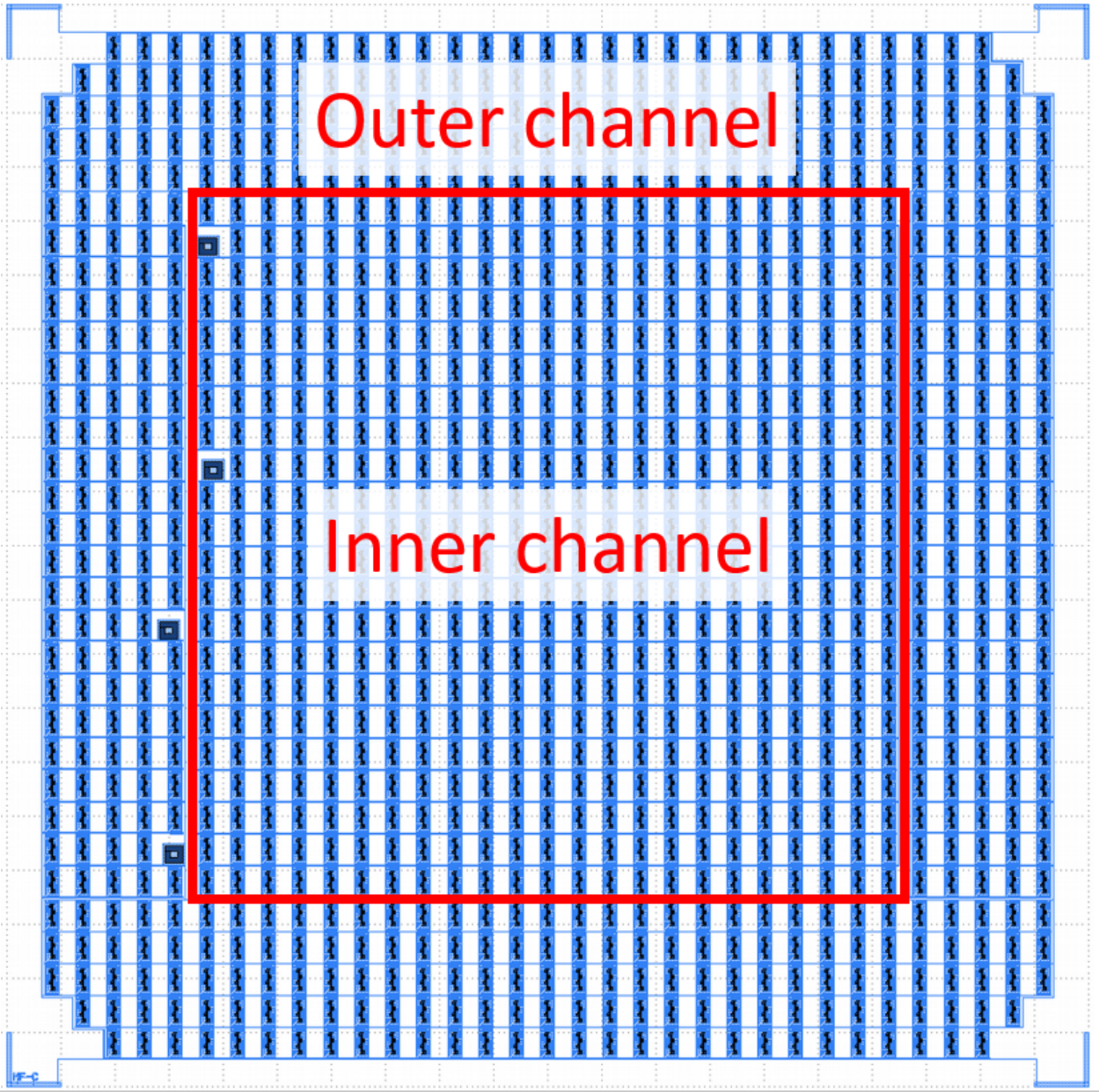}
    \caption{Left panel: Photograph of the HVeV detector surrounded by the black frame. The electrical connections on the printed circuit board (PCB) are visible on the left side of the detector, while a second detector that was not used in this work can be seen on the right side. Right panel: Scheme of the QET pattern covering the detector top surface. The two phonon channels, comprised of parallel TES arrays, are divided by the red line; the four filled squares on the left side are the contacts used for wire bonding to the PCB.}
    \label{fig:HVeV}
\end{figure}
A photograph of the detector mounted in its light-tight copper holder is shown in Fig.~\ref{fig:HVeV} (left).
The detector consists of an instrumented 1\,cm\,$\times$\,1\,cm\,$\times$\,0.4\,cm Si chip clamped between two printed circuit boards (PCBs). The detector top surface is instrumented with two QET arrays. Each of these arrays forms a single readout channel. The outer channel frames the inner one to provide event-position information. The two sensors are arranged to provide equal-area coverage of the inner and outer portion of the detector as shown in the right side of Fig.~\ref{fig:HVeV}. An electric field of $0-625$~V/cm was set across the detector by maintaining the QET face at ground and biasing an Al grid on the detector face opposed to the instrumented one.

\begin{figure}[h]
    \centering
    \includegraphics[width=1\columnwidth]{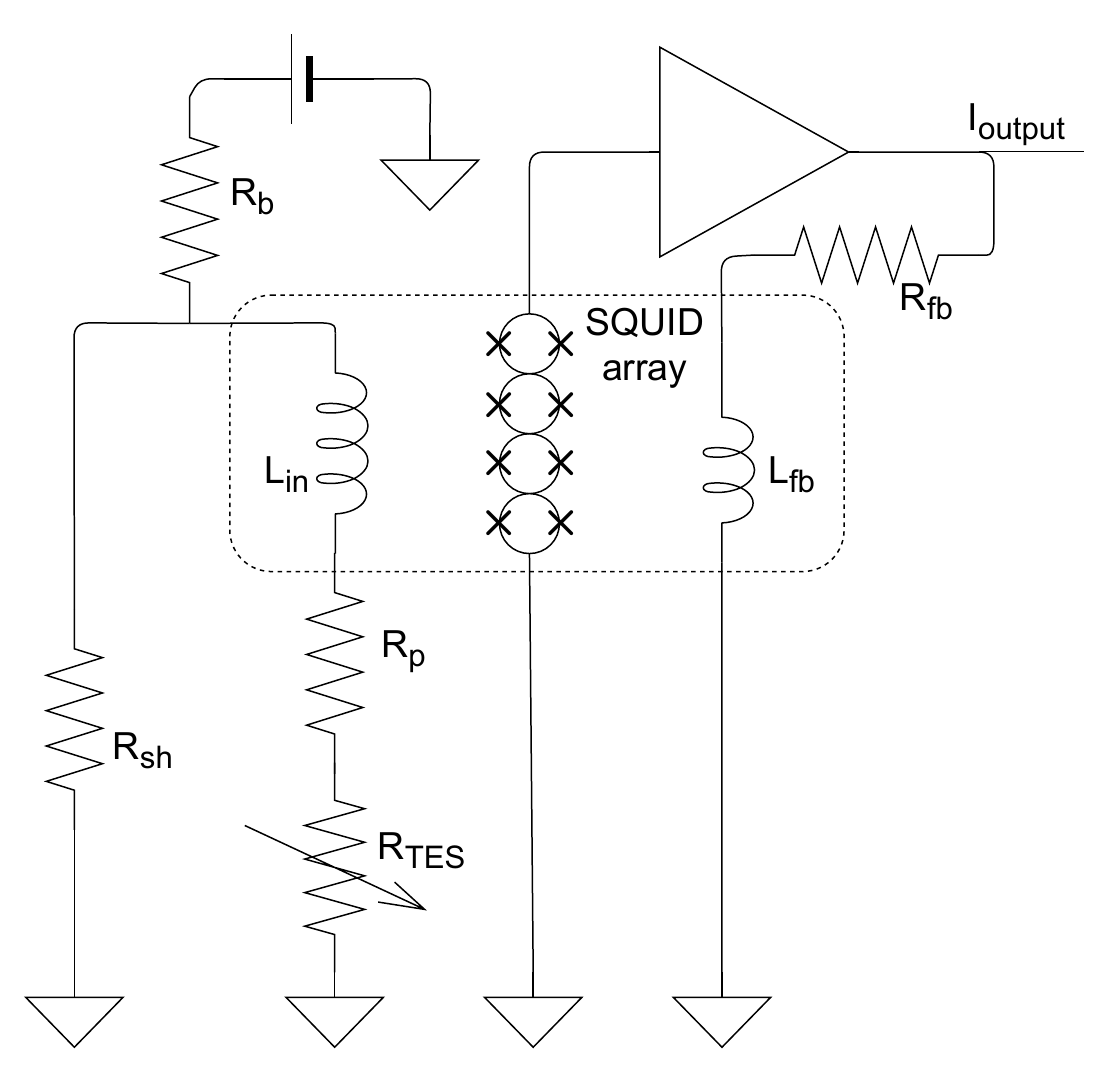}
    \caption{Scheme of the readout circuit for one TES channel represented by $R_{TES}$. The dashed box indicates the components present on the SQUID chip. The parasitic resistance on the TES branch is represented by $R_p$, while is included in the evaluation of the shunt resistor $R_{sh}$ itself in the shunt resistor branch. 
    The inductance of the input coil and the feedback coil are referred as $L_{in}$ and $L_{fb}$ respectively}
    \label{fig:circuit}
\end{figure}

For this work, we generally operated the TESs between 30\% and 45\% of R$_n$, slightly below the midpoint of the transition regions in order to maximize the dynamic range of the detector. The signal currents were read out using Superconducting Quantum Interference Device (SQUID) amplifiers operated in a flux-locked feedback loop. 
Figure~\ref{fig:circuit} shows a scheme of the readout circuit, which will be also referred later in the following section.
The output signals were digitized with a 16-bit National Instruments PCIe-6374 DAQ, with a 1.51~MHz sampling frequency. The signals were digitized continuously with  triggering and pulse analysis performed offline.

We operated this detector at $\sim 50$~mK in a Vericold Adiabatic Demagnetization Refrigerator (ADR)~\cite{Hong:2020} both at Northwestern University (Evanston, IL) and at the Triangle Universities Nuclear Laboratory (Durham, NC). During each of the $\sim$1-month-long data taking periods, the ADR was cooled down daily from 4~K to the working temperature. The detector's working point was set everyday and daily calibrations were required.  

The low-energy calibration was performed using pulses from a 635-nm laser. The laser was operated at room temperature with an optical fiber passing from room temperature to the detector at $\sim50$~mK. We used an inline infrared filter at 1.4~K to suppress black-body radiation from room-temperature and the warmer ADR stages (for more details see Ref.~\cite{Hong:2020}).
The laser diode was driven with a pulsed current source. The laser-on time was set to 500~ns, which is small compared with the $20~\mu\mathrm{s}$ rise time of this detector. We varied the current to alter the laser intensity, which produces Poisson-distributed photon bursts with a measured average number of photons per pulse, $\lambda$, ranging from 0.2 to 20. For each laser pulse, we generated a TTL-like digital signal that was recorded in the National Instruments DAQ and used to synchronize the laser pulses in the offline analysis.

The calibration was extended to higher energies using external $^{55}$Fe and $^{57}$Co sources. The ADR has a Be window facing the experimental volume which permitted penetration of soft X-rays to the detector. A set of aluminized Mylar layers interposed between the Be window and the detector provided black-body radiation shielding while ensuring minimal X-ray attenuation.

After the initial detector operations, we moved this program to a Cryoconcept dry dilution refrigerator hosted at the ``Northwestern EXperimental Underground Site'' (NEXUS)~\cite{Battaglieri:2017} that allowed us to characterize an NF-C detector at colder temperature (10~mK), complementing the data already acquired with the ADR.
NEXUS is a shallow facility, located $\sim$100~m underground in the MINOS near-detector experimental hall at the Fermi National Accelerator Laboratory (FNAL). It is maintained and operated through a collaboration between Northwestern University and FNAL. The dilution-unit-based cryostat allowed more stable operation at lower temperatures compared to the ADR. At NEXUS, the detector was controlled and read out using custom detector control and readout cards developed for SuperCDMS~\cite{Hansen:2010}. These interfaced with the SQUID readout system and digitized the phonon signals at 625~kHz.

\section{QET Characterization}\label{sec:char}
\begin{table*}[t]
    \centering
    \caption{Detector design parameters for the QP.4 prototype device (Ref.~\cite{Hong:2020}) and the NF-C device described in this paper. Numbers for QP.4 are measured values from the previous reference, while numbers for NF-C are model predictions based on changes in the detector mask design. Both detectors have two channels, an inner grid of QETs surrounded by an outer frame of QETs. The outer channel in the QP.4 device had fewer QETs than the inner channel. Where multiple numbers are presented, the first/second number in the column is for the outer/inner channel on that device. The NEXUS analysis has not yet been extended to measure energy efficiency, as we are trying to improve the precision of the measurement, but the resolution implies it will be comparable to the efficiency found in the ADR. Similarly, we were unable to measure complex impedance in the ADR, so a precise estimation of NEP was not possible.}
    \begin{tabular}{|c|c|c|c|c|c|c|c|}
        \hline
        Parameter & Description & Unit & QP.4~\cite{Hong:2020} & \multicolumn{4}{c|}{NF-C} \\
        \hline
        $T_b$ & Bath Temperature & mK & 50&\multicolumn{2}{c|}{$\sim$50} & \multicolumn{2}{c|}{$\sim$10} \\
        \hline
         & Design/Cryostat & & ADR&Design&ADR &Design & NEXUS\\
        \hline
        $A_{det}$ & Detector Area & $\mathrm{cm}^2$ & \multicolumn{5}{c|}{1} \\
        \hline
        $T_c$ & TES Critical Temperature & mK & \multicolumn{1}{c|}{$\sim$65} & \multicolumn{4}{c|}{60 -- 65} \\
        $\eta$ & Detector Thickness & mm & 1 & \multicolumn{4}{c|}{4} \\
        $m_{det}$ & Detector Mass & g & 0.24 & \multicolumn{4}{c|}{0.96} \\
        $N_\mathrm{QET}$ & QETs per Channel & - & 170/300 & \multicolumn{4}{c|}{504/536} \\
        $l_\tes$ & TES length & $\mu$m & 100 & \multicolumn{4}{c|}{150} \\
        $v_\tes$ & TES Volume (Per Channel) & $\mu$m$^3$ & 1360/2400 & \multicolumn{4}{c|}{$7.39\times10^3$} \\
        $\zeta_\tes$ & Fraction of W in TES & - & 0.5 & \multicolumn{4}{c|}{0.5} \\
        $v_{eff}$ & Effective W Volume & $\mu$m$^3$ & 2720/4800 & \multicolumn{4}{c|}{$1.48\times10^4$} \\
        $l_{fin}$ & Al Fin Length & $\mu$m & 125 & \multicolumn{4}{c|}{60} \\
        \hline
        $\rho/\eta_\tes$ & Resistivity/Thickness Ratio & $\Omega$ & 2.88 & 2.88 & 3.0~$\pm$~0.3 & 2.88 & 3.0~$\pm$~0.3\\
        \multirow{2}{*}{$R_n$} &  Normal Resistance (Inner) &  \multirow{2}{*}{\mohm{}} & 400 & 350 & 332 -- 396 & 350 & 332 -- 396 \\
        & Normal Resistance (Outer) & & 700 & 350 & 311 -- 371 & 350 & 311 -- 371\\
        \hline
        $P_{Chan}$ & Bias Power (Channel) & pW & 1.2/2 & 4.6 -- 8.4 & 4.0~$\pm$~0.6 & 7.5 -- 11.5 & 6 -- 7.5 \\
        $P_\tes$ & Bias Power (Per TES) & fW & $\sim$7 & 8.8 -- 16.0 & 7.6~$\pm$~1 & 14.3 -- 21.9 &  13~$\pm$~2 \\
        $G_{Chan}$ & Thermal Conductance (Channel) & pW/K & 120/200 & 640 -- 880 & 350 -- 650 & 640 -- 880 &  460 -- 625 \\
        $G_\tes$ & Thermal Conductance (Per TES) & fW/K & 225/375 & 1220 -- 1680 & 660 -- 1250 & 1220 -- 1680 &  880 -- 1190 \\
        $\Sigma_{eph}$ & TES Electron-Phonon Coupling Constant & GW/($\mathrm{K}^5\cdot \mathrm{m}^3$) & 0.47 & 0.47 & 0.27 -- 0.67 & 0.47 &  0.35 -- 0.65 \\
        \hline
        $\tau_{\mathrm{BW}}$ & Pulse Fall time & $\mu$s & $\sim100$ & 70 -- 160 & $\sim80$ & 55 -- 100 & $\sim$30\\
        $\epsilon$ & Energy Efficiency & - & $\gtrsim$22\% & 27\% & $\gtrsim$29\% & 27\% & - \\
        $\sigma$ & Resolution & eV & 3.0$\pm$0.5 & 2.3 -- 2.4 & 2.65$\pm$0.02 & 1.8 -- 2.1 & $\sim$2.9~eV\\
        $S_p$ & NEP (Channel) & aW/$\sqrt{\mathrm{Hz}}$ & 5.3 & 11 -- 14 & - & 11 -- 14 & 10 \\
        $S_p$ & NEP (Per TES) & zW/$\sqrt{\mathrm{Hz}}$ & 0.23 & 0.5 -- 0.6 & - & 0.5 -- 0.6 & 0.4 \\
        \hline
    \end{tabular}
    \label{tab:devicePerformance}
\end{table*}

\begin{figure*}[t]
    \centering
    \raisebox{-0.5\height}{\includegraphics[width=0.45\textwidth]{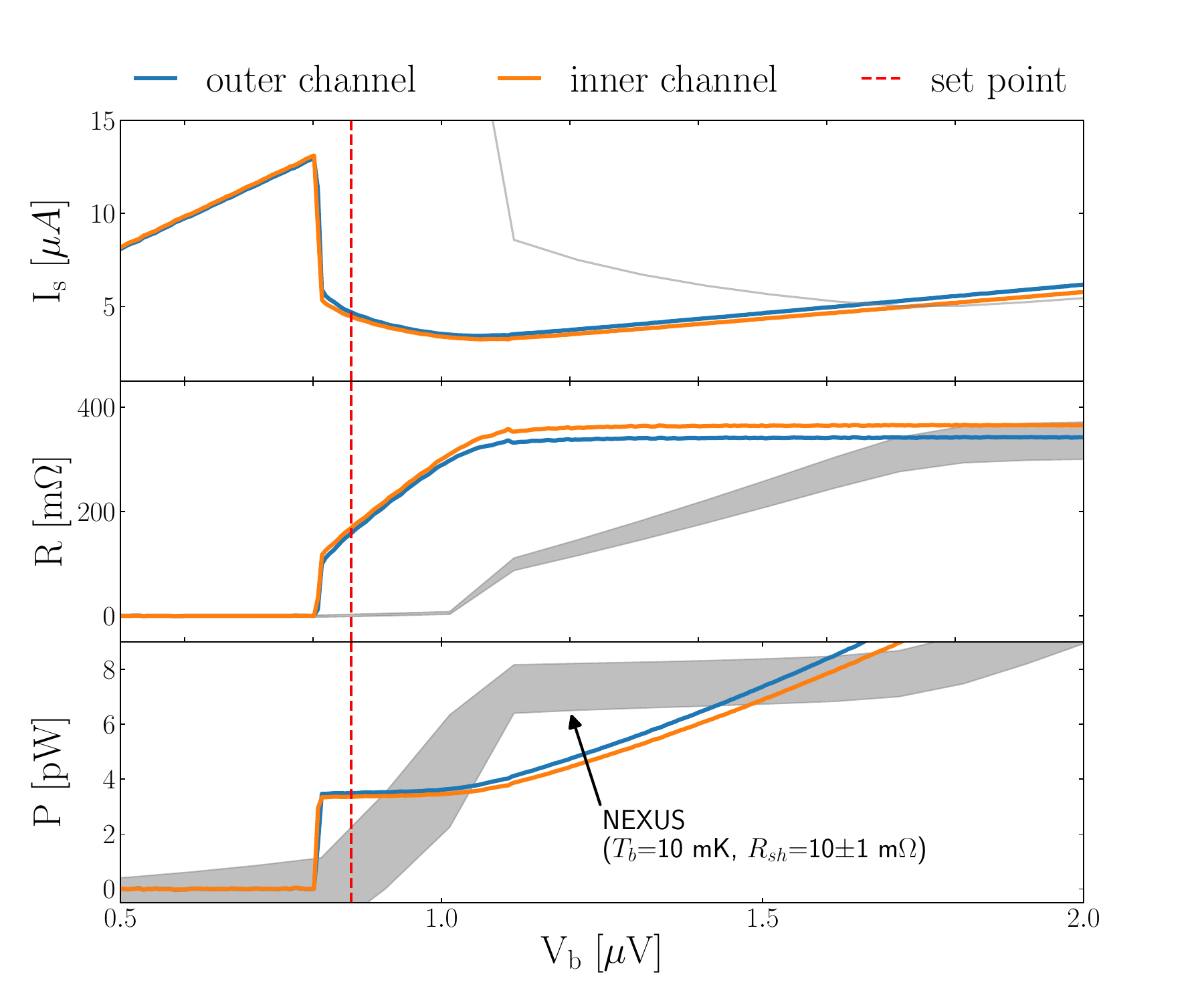}}
    \raisebox{-0.5\height}{\includegraphics[width=0.50\textwidth]{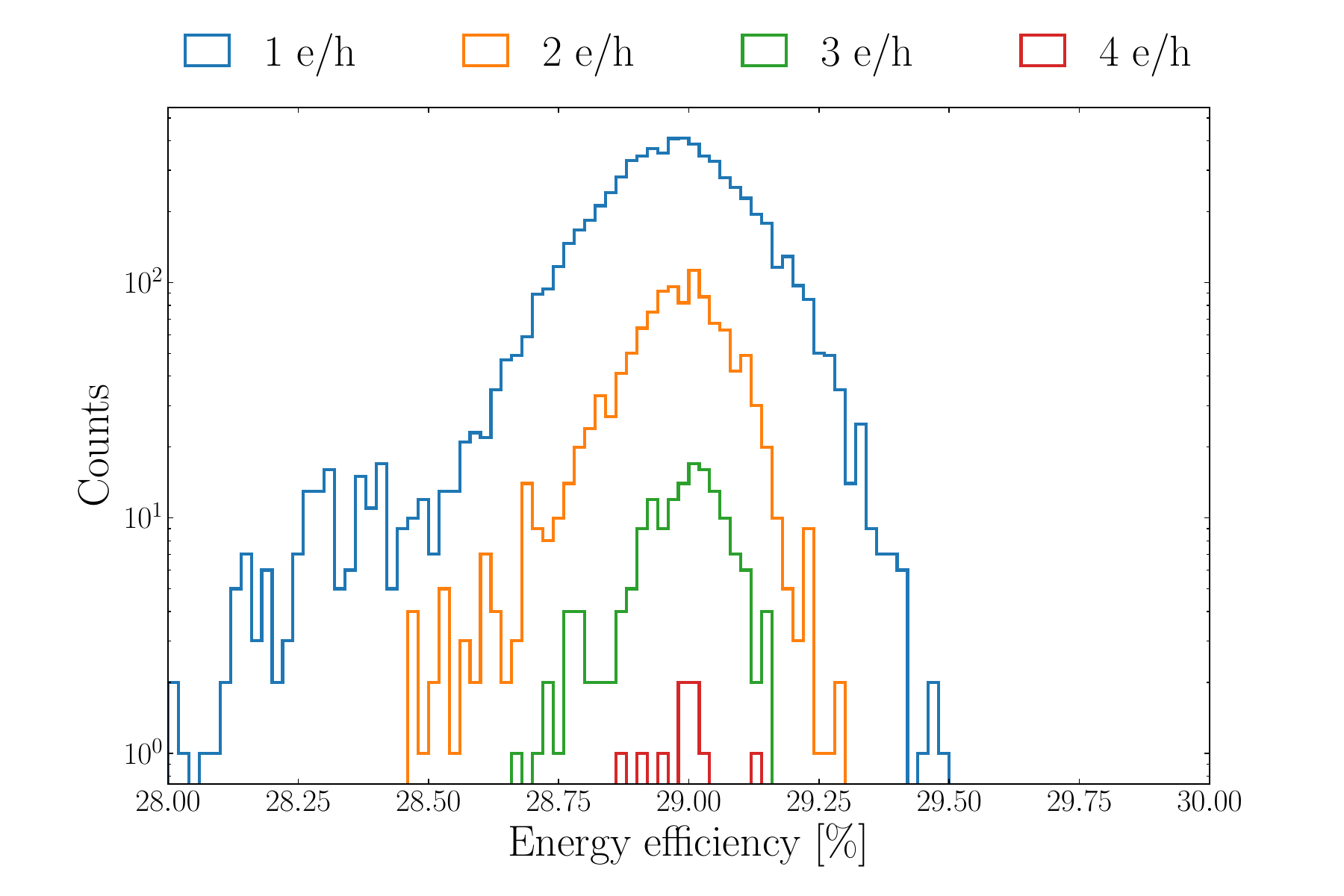}}
    \caption{Left panel: TES signal current, resistance and power curves as a function of bias voltage $V_b=I_bR_{sh}$ for the two channels of NF-C operated at 50~mK in the Vericold ADR. 
    In addition, the same variables are measured with a single channel operated at 10~mK in the NEXUS cryostat. The thickness of all the curves represents the associated systematic uncertainty band. The band is larger in the NEXUS data because of its lower value of shunt resistor ($R_{sh}\approx 10$~\mohm) and associated larger systematics uncertainty compared to the 50~m$\Omega$ shunt resistor used in the ADR}. 
    The detector was operated at 45\% of the normal resistance value when operated in the ADR. The detector was operated at 30\% of the normal resistance during the measurements at NEXUS. Right panel: Reconstructed energy efficiency for different numbers of electron-hole pairs, see the text for details on the circuit parameter used.
    \label{fig:IV}
\end{figure*} 

To validate the detector model, we measured various QET array properties such as bias power, energy efficiency and power noise. These measurements are key to understanding any differences between estimated and measured energy resolution. The measurements presented in this section are in good agreement with those predicted by our detector model, as shown in Table~\ref{tab:devicePerformance}.

\subsection{Resistance and Bias Power}

Basic TES parameters can be evaluated by scanning through values of the TES bias voltage, $V_b$, and measuring the DC current response from the QET channel. In the first row of Fig.~\ref{fig:IV} (left) we show the variation of signal current $I_s$ with bias voltage for both channels of NF-C. In the second row, we have calculated the inferred channel resistance~\cite{Irwin:2005}, $R$, and in the final row the Joule power produced, $P=I_s^2R$. We note that these scans were performed on the two channels simultaneously. 

The bias power is affected by other heating effects like operating both channels simultaneously or changing the bath temperature. 
The reciprocal over-heating of the two channels lowers the required joule heating to stay in transition.
Also, a higher bath temperature lowers the bias power needed to stay in transition.
We note that the bias power is lower than was predicted by the NF-C model in the ADR measurement.
For this reason, we repeated the measurement at NEXUS operating only one channel and at a lower bath temperature in Fig.~\ref{fig:IV} (left).
The contribution of reciprocal heating was observed to be 1 to 2~pW at NEXUS. The bias power measured at NEXUS is then used in Sec.~\ref{sec:noisemod} for the noise modeling.

\subsection{Energy Efficiency}
For a TES in strong feedback~\cite{Irwin:2005}, the phonon energy absorbed by a TES can be inferred from the change in signal current and circuit parameters as
\begin{equation}
\begin{split}
    E_{abs}&\approx\left(1-2\frac{R_{\ell}}{R_{\ell}+R_0}\right)I_bR_{sh}\int\delta I_s(t) dt\\
    &+ R_{\ell}\int{\delta I_s}^2(t)dt
\end{split}
\end{equation}
where $R_{\ell}=R_{sh}+R_{p}$ is the total resistance (shunt and parasitic) in the TES bias loop~\cite{Irwin:2005} apart from the TES, $R_0$ is the TES operating resistance, and $I_b=V_b/R_{sh}$ is the TES bias current. Here we have defined $\delta I_s(t)=I_0-I_s(t)>0$ as the change in signal current during a phonon pulse relative to the quiescent value, $I_0$.
This absorbed energy can be compared to the calibrated total phonon energy to define the detector's energy efficiency, $\epsilon = E_{abs}/E_{ph}$.

The efficiency was evaluated using a laser calibration dataset with a mean number of photons per pulse $\lambda\sim0.3$, the detector operated at $V_{\mathrm{NTL}}=100$ V, and the cryostat temperature stable at $50.00 \pm 0.01$ mK. Data selection criteria were applied to select pulses which were coincident with the laser trigger signal, had energy above the noise threshold, and had a stable baseline signal before the pulse.

Figure~\ref{fig:IV} (right) shows the energy collection efficiency that was calculated for individual phonon pulses using a particular set of circuit parameters.
For this figure, we selected the most conservative set of assumptions to obtain a lower estimate of the energy collection efficiency of $\epsilon\gtrsim29\%$. As reported in Table~\ref{tab:devicePerformance} and detailed in Appendix~\ref{app:eff}, this is compatible with design expectations. 
The current measurement is dominated by the systematic uncertainties in TES circuit parameters (e.g.~R$_p$ and R$_0$); future measurements will include more precise characterization of these components to place tighter constraints on this value.

\subsection{Noise Modeling}\label{sec:noisemod}
The resolution model for a QET described in Section~\ref{sec:design} relies on the assumption that the QET noise is dominated only by thermal fluctuations across the thermal conductance $G$ between the TES and the crystal. In reality, the bias circuit has its own intrinsic noise from both passive components and the SQUID current amplifier. Optimization of the detector normal resistance takes these expected contributions into account to ensure that the TES is dominated by its own quantum noise. Modeling the current noise, and converting to Noise Equivalent Power (NEP), allows us to compare the intrinsic power noise of the QET to that expected by the resolution model. The NEP for a generic thermal detector with thermal conductance $G$ at temperature $T=T_c$ is~\cite{Irwin:2005}
\begin{equation}
    NEP = \sqrt{4k_b T_c^2 G}
\end{equation}
and thus we can compare the noise power inferred from the current noise to the expectation from the measured bias power and transition temperature, which predicts the magnitude of the thermal fluctuation noise and is expected to be flat in NEP. The NEP expected for these detectors is summarized in Table~\ref{tab:devicePerformance}.

In order to validate this noise model and demonstrate that this detector achieved near quantum-limited noise, we employed the TES bias circuit noise model described in past work (see e.g. Refs.~\cite{Fink:2020,Kurinsky:2018,Pyle:2012,Irwin:2005}). 
Due to the less constrained input inductance and parasitic resistance parameters on the ADR electronics circuit, we carried out a dedicated measurement at NEXUS.
We characterized the noise inherent to the SQUID bias circuit using a SQUID with the TES coil disconnected. We then fit the contribution of passive noise to the total transition noise by adjusting the effective noise temperature of the fit to jointly match the noise in the normal and superconducting states.
We also measured the complex impedance of the TES both with a square wave impulse and swept sine wave measurements to characterize the TES thermal poles, with results summarized in Table~\ref{tab:noiseParams}. 
The superconducting noise combined with complex impedance measurements constrained the inductance in the loop. 
In addition, we were able to extract estimates of TES response characteristics in Table~\ref{tab:noiseParams}  (similar to the method used in Ref.~\cite{Fink:2020}) to constrain the TES power to current response. The measurements of bias power in the lower temperature environment in NEXUS allows us to bound thermal conductance and better constrain the parameters in Table~\ref{tab:devicePerformance}.

\begin{figure}[t]
    \centering
    \includegraphics[width=\columnwidth]{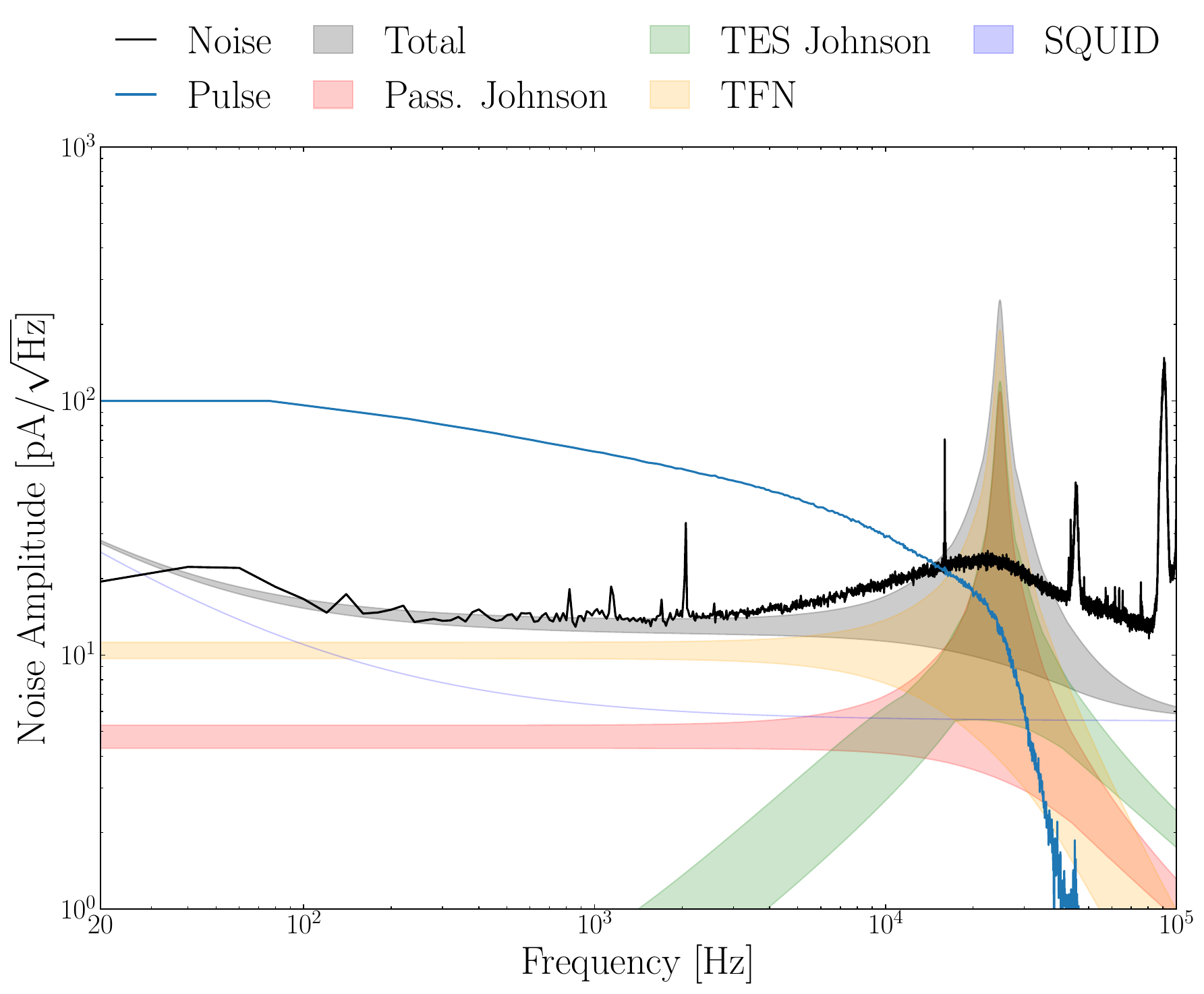}
    \includegraphics[width=\columnwidth]{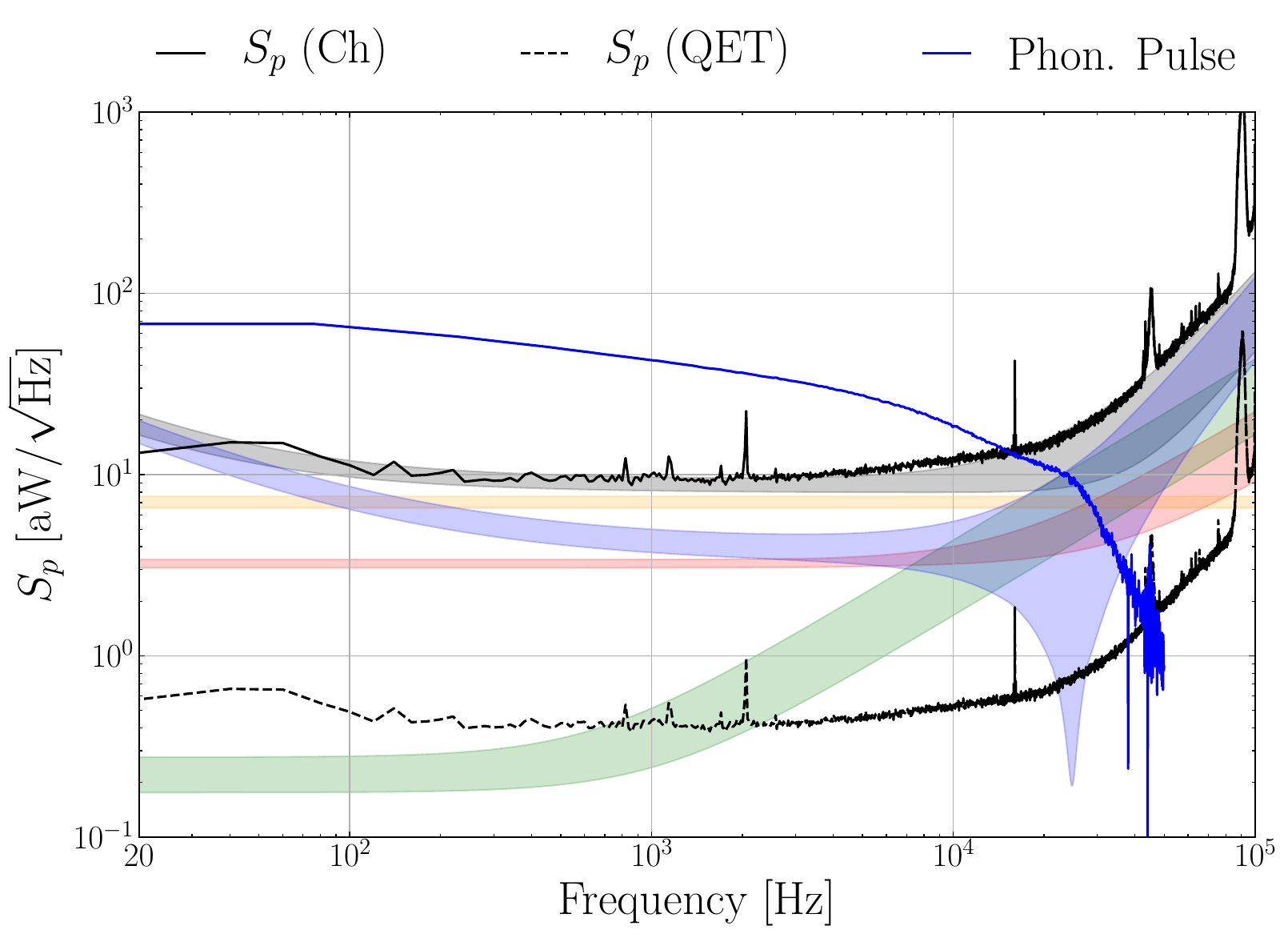}
    \caption{Top panel: Current noise for NF-C run in NEXUS (black) compared to the best-fit model informed by complex impedance measurements taken in the same facility, highlighting the dominance of thermal fluctuation noise (TFN) assumed for the detector modeling. The pulse shape found by averaging pulses near threshold is also shown. The pulse shape is scaled arbitrarily relative to noise to better visualize atop the noise. Bottom panel: Power noise inferred from noise modeling, computed by dividing the current noise by the power to current transfer function derived from the complex impedance measurement~\cite{Fink:2020}. The total QET channel power noise, as well as the noise per individual QET cell are shown in black, compared to the pulse shape (blue) in power space. In both cases, the closed-loop SQUID gain begins to drop around 50~kHz, where the phonon pulse is cutoff. This also artificially broadens the electrothermal oscillation peak at $\sim$25~kHz.}
    \label{fig:noise}
\end{figure}

The measured current noise for a single QET channel at the operating bias point of $R_0/R_n=0.43$ is shown in Fig.~\ref{fig:noise} (top), along with the model incorporating systematic uncertainties, demonstrating that the TES response is dominated by the quantum (thermal fluctuation) noise. At high frequency, the signal to noise was degraded by electrothermal oscillation due to the high inductance of the readout system ($\sim$800~nH), which impacts both the QET pulse and the noise. Dividing out the electrical response of the TES bias loop using complex impedance measurements gives the estimates for noise equivalent power in Fig.~\ref{fig:noise} (bottom). With around 525~QETs/channel, we obtain a total power noise of $\sim$10~aW/$\sqrt{Hz}$, which is equivalent to 500~zW/$\sqrt{Hz}$ per individual QET cell. This is consistent with the NEP used to estimate TES resolution in Eq.~\ref{eq:sigma}.

The large error bands in the noise model come from the same source of systematic uncertainty as for energy efficiency, namely the uncertainty in overall resistance scale. This becomes a systematic uncertainty on bias power, leading to a large range in the measurement for $G$, but is also degenerate with measurements of inductance. In addition, some uncertainty comes from the limited bandwidth of the measurement technique used for the data taken in this run. Future measurements will further constrain QET properties by carefully calibrating out these uncertainties and by improving the precision of the complex impedance characterization. In particular, four-wire measurements of the detector $R_n$ will reduce the large systematic uncertainties on the resistance scale, which dominates the uncertainty of all measurements described in this paper.

\begin{table}[t]
\caption{TES bias circuit parameters measured at NEXUS, which were extracted from complex impedance measurements used to fit TES noise in Fig.~\ref{fig:noise} (parameters refer to the definitions employed in Ref.~\cite{Fink:2020}). Fall time and feedback gain in electrothermal feedback (ETF) parameterize the effect of the voltage bias feedback on the TES response. The reported uncertainties are dominated by the systematic uncertainty on the shunt resistor and affected by the limited bandwidth of the readout circuit. Distortions in the driving signal above $10$~kHz required a correction for finite bandwidth in the bias circuit. Higher precision measurements will better constrain these parameters as a function of bias point and base temperature in future work.}
\begin{tabular}{|c|c|c|}
\hline
    Parameter & Description & Value \\
    \hline
    L & Inductance & 850$\pm$50~nH\\
    $R_{sh}$ & Shunt Resistance & 8$\pm$1~m$\Omega$ \\
    $R_{p}$ & Parasitic Resistance & 19$\pm$2~m$\Omega$ \\
    $R_0/R_n$ & Bias Point & 0.43$\pm$0.01 \\
    $R_{0}$ & TES Resistance & 125$\pm$25~m$\Omega$\\
    $|\tau_{ETF}|$ & ETF Fall time & 7--8$\mu$s \\
    $\tau_0$ & Thermal Fall time & 200$\pm$50~$\mu$s \\
    $\mathcal{L}$ & ETF Gain & 30$\pm$5 \\
    $\beta$ & Current Response & 0.2-0.3 \\
\hline
\end{tabular}
\label{tab:noiseParams}
\end{table}

\section{Event Reconstruction}\label{sec:recon}

\begin{figure*}[t]
    \centering
    \includegraphics[width=0.50\textwidth]{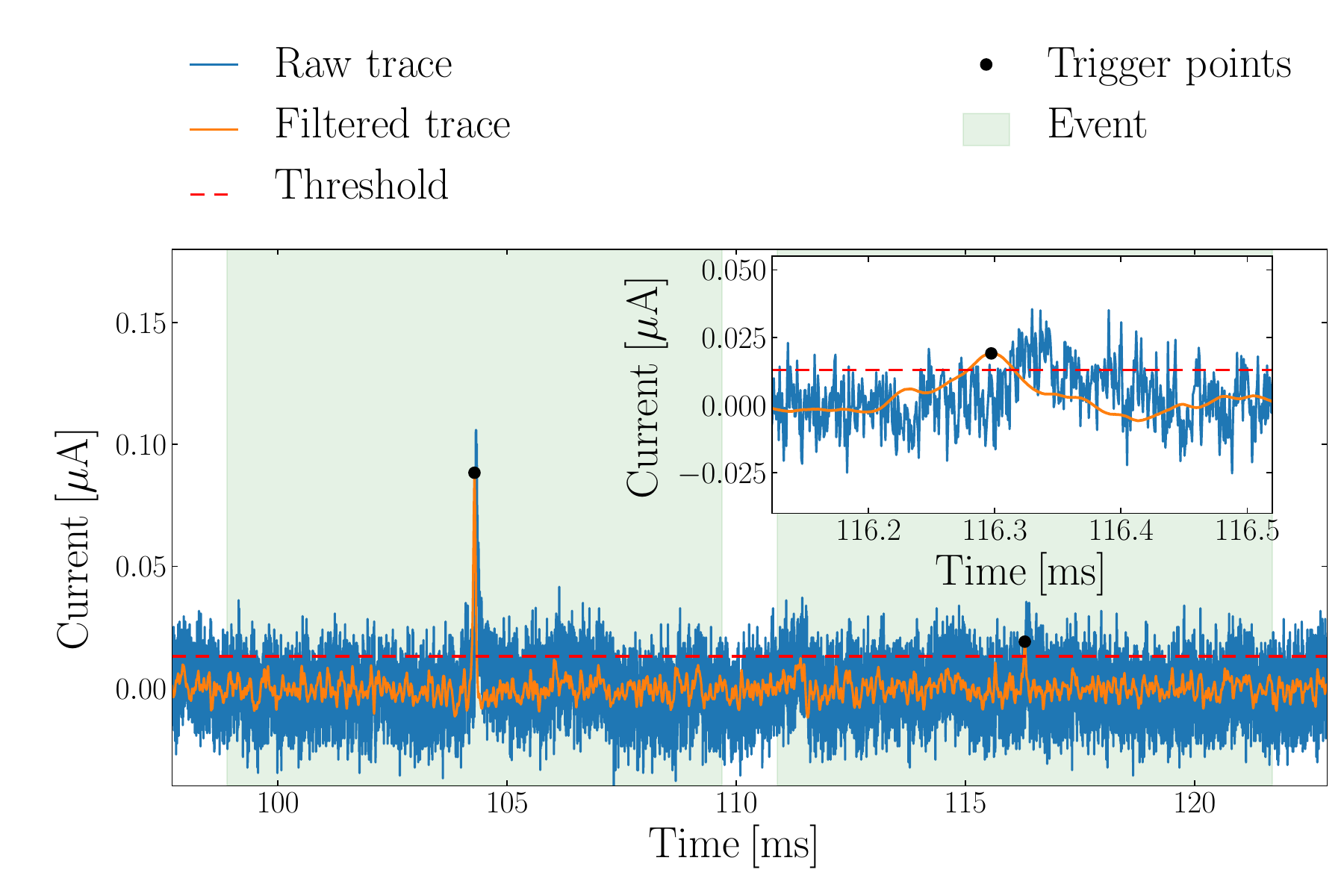}
    \includegraphics[width=0.49\textwidth]{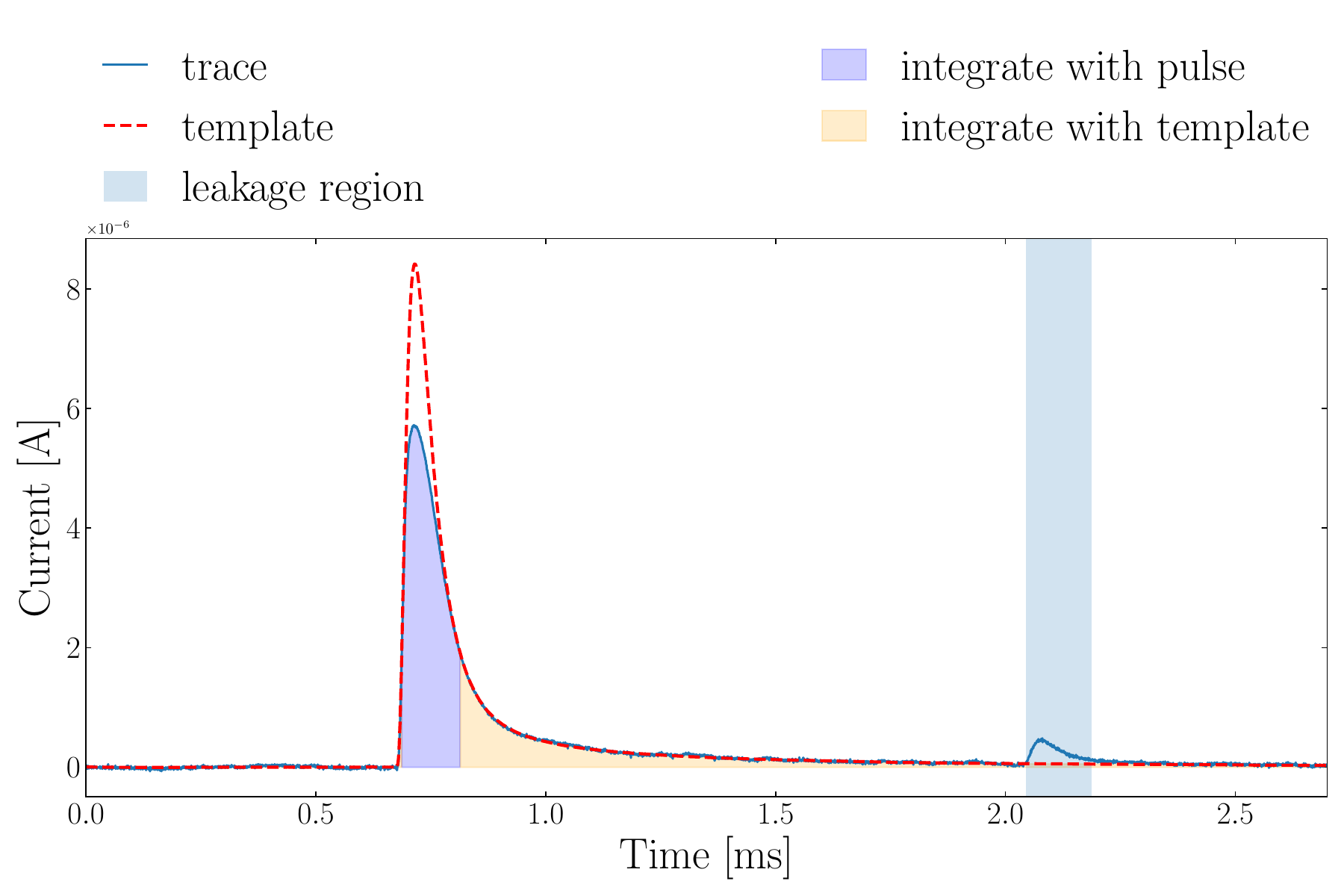}
    \caption{Left panel: Illustration of the triggering algorithm. A raw trace is shown in blue, while the optimum filtered trace is shown in orange. The dashed line is the trigger threshold. The green vertical bands show the trace regions selected by the triggering algorithms as events.  Inset: a zoomed in piece of the trace around a small pulse. The filtered trace reaches its maximum value at the pulse onset.
    Right panel: Illustration of the matched-filter-based energy estimator, MF integral, used to enlarge the energy range of the detector. The primary event is shown around 0.7~ms. 
    The area highlighted in violet corresponds to the part integrated using the pulse itself. The area highlighted in yellow is integrated as the area below the red template, which is fit to the pulse tail in the yellow range.
    The pileup of the ``leakage'' pulse is identified through a threshold trigger and excluded from the tail fit to minimize its impact on the energy estimate of the primary pulse.}
    \label{fig:trigger_example}
\end{figure*}

Data were acquired as a continuous time-stream and were processed offline. Different trigger and energy estimators were used depending on the purpose of the analysis.
The optimization of energy resolution and threshold are fundamental for a nuclear-recoil dark-matter search. A time-domain Optimum Filter (OF) trigger was used to reach the lowest threshold while maintaining a relatively low trigger rate. The same filtering technique was used for energy reconstruction to optimize the resolution at low energies.
In opposition, the ionization-yield measurement required a larger dynamic range (between tens of eV and tens of keV) but the constraints on the threshold and energy resolution were looser.
The data were triggered with a higher threshold using the matched-filter-based trigger algorithm described in Ref.~\cite{Amaral:2020}.
An integral-based energy estimator was used to increase the dynamic range. The following two subsections describe the OF trigger algorithm and the integral energy estimator used in this paper.

\subsection{Optimum Filter Trigger}
The OF is a minimum variance estimator of the amplitude of a pulse, with a known shape, in the presence of stationary noise, as described in Refs.~\cite{Gatti:1986,Fowler:2015}. 
In the current work, we used the OF in order to trigger with lowest achievable threshold similar to Ref.~\cite{Domizio:2011}.

Several laser data sets---approximately equally distributed in time over operations---were used to construct the pulse template. Pulses coincident with the laser signal were collected by triggering on a digital trigger signal from the laser driver. These events were averaged to produce the pulse template. The length of the template was optimized empirically to get the best energy resolution on the laser data and set to 16384 samples (10.8~ms). The noise PSD was evaluated by collecting noise traces of the same length, using a random trigger and applying a pulse rejection algorithm to select pulse-free traces. This algorithm makes an iterative rejection of outlier events based on mean, range, slope and skewness of the traces. The outlier rejection procedure iteratively removes events furthest from the median of the distribution until the skewness of the distribution is less than 0.05.

After constructing both the noise PSD and the pulse template, a data stream was filtered and a threshold trigger was applied to the filtered trace. A peak search window was defined spanning 8192 samples after the crossing point for each threshold crossing occurrence. The trigger point was then adjusted to the point where the filtered trace reaches its maximum value within the peak search window. A snippet of the raw trace within $\pm$8192 samples around the adjusted trigger point defines a triggered event which then undergoes further processing, where various event parameters are being evaluated, such as the template fit chi-square, the integral of the trace, the mean value of the pre-pulse region. An example of the trigger algorithm applied to a pulse can be seen in Fig.~\ref{fig:trigger_example} (left).

\subsection{Integral-based energy estimators}
At energies below $\sim$1~keV, the amplitude provided by the OF was used as an energy estimator to get the best possible resolution. However, TES saturation effects at higher energies distort the pulse shape, producing a large non-linear response and eventually saturating the OF estimator itself. 

A hybrid of a pulse integral and a template fit was used to increase the dynamic range for high-energy analyses. The goal was to get the best estimation of the area of the pulse with a direct integral for the part where the pulse amplitude is high but distorted by saturation effects, while using a fit to a pulse template to estimate the area where the detector behaves linearly but the signal to noise is low.
We integrated the region where the pulse is above 2~$\mathrm{\mu A}$. The rest of the pulse was fit to a pulse template and then integrated from the 2~$\mathrm{\mu A}$ crossing to the end of the pulse window. A 2.7-ms-length window was used, where the pre-pulse corresponds to 0.7~ms.
The choice of a shorter trace with respect to the OF was dictated by a looser requirement for the energy resolution, which was in any case limited by the integral-based energy estimator.
The $2~\mathrm{\mu A}$ threshold was chosen as the level where the signal level is much higher than the noise level before the onset of heavy saturation. The resulting estimator is therefore functionally a hybrid of pure integration and a matched filter (MF), integrating the high signal to noise region of the pulse directly and using the MF to estimate the contribution of the tail to the total pulse energy to reduce integrated noise. In addition, if there is a pileup pulse present between the 2~$\mathrm{\mu A}$ crossing and the end of the trace, the pileup-pulse region is excluded from the fit. This exclusion region is defined as 10~$\mathrm{\mu s}$ before the pileup-pulse trigger to 130~$\mathrm{\mu s}$ after it, which is effective for preventing the dominant source of pileup pulses (single-electron-hole-pair leakage) from significantly affecting the fit. This energy estimator is referred to as the \textit{MF integral} in the rest of the paper. Figure~\ref{fig:trigger_example} (right) illustrates the described procedure.

\section{Detector Performance}\label{sec:perf}

The results of the Section~\ref{sec:char} suggest that the parameters that feed into the energy resolution estimate match expectation, and thus we should find the energy resolution to be close to the design expectation. In this section, we report a measured baseline resolution\footnote{We refer to baseline resolution as the detector energy resolution when no pulses are recorded.} comparable to the design value---2.65(2)~eV compared to $2.3-2.4$~eV expectation---and explore the small signal response. We then discuss the performance of the MF integral estimator and the calibration used to extend the energy scale to 120~keV, corresponding to an effective dynamic range of 4 orders of magnitude. We also analyze the difference between the 0~V and high-voltage (HV) energy scales and potential causes for the small differences that are observed.

\subsection{Small Signal Response}

Calibration of the low-energy region (below $\sim$1~keV) is performed with laser data sets as described previously.
The single-charge resolution leads to discrete peaks in the spectrum corresponding to quantized charge excitation. This produces a set of well defined lines of known energy that can be used for calibration.
\begin{figure}[ht]
    \centering
    \includegraphics[width=\columnwidth]{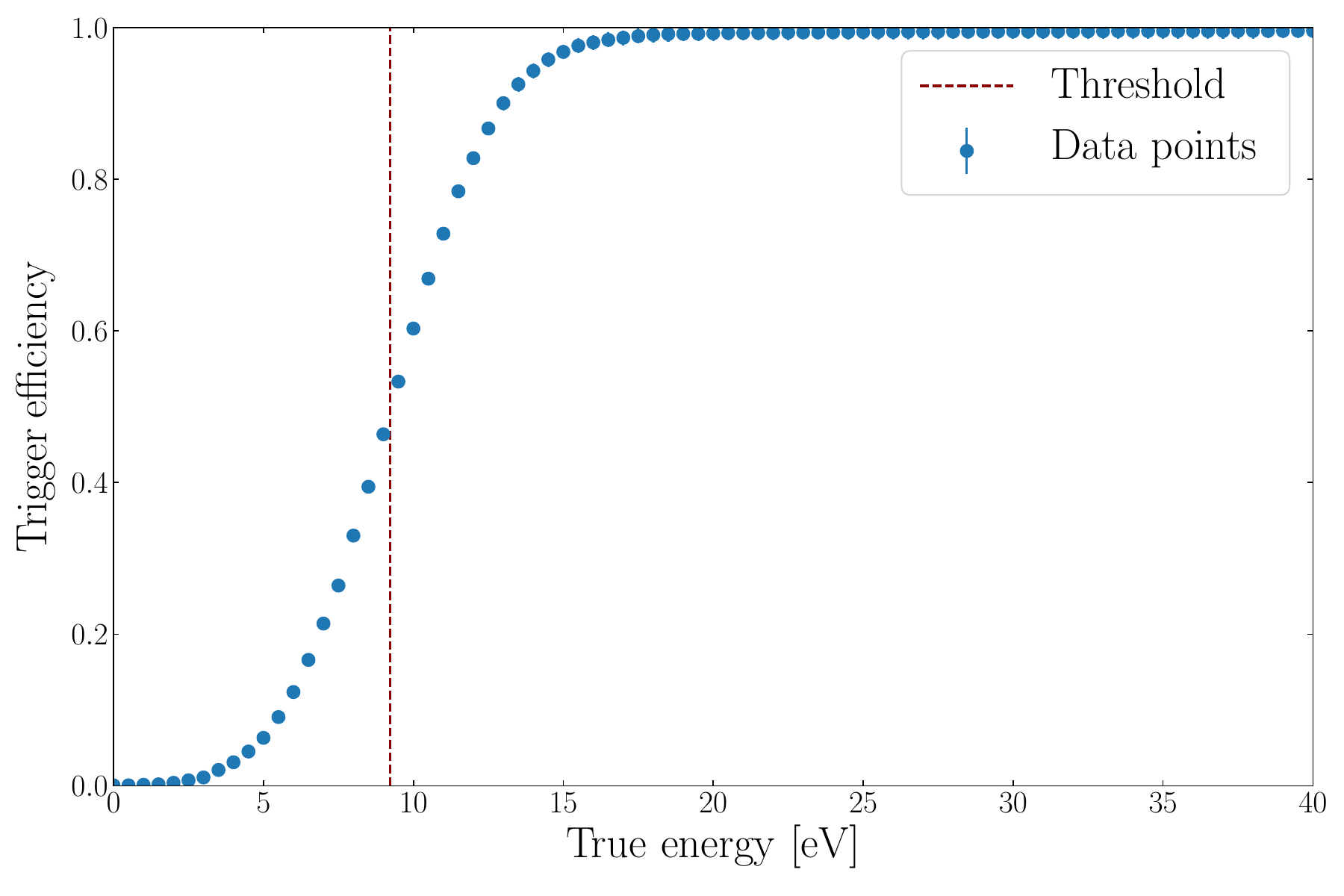}
    \caption{Trigger efficiency, measured by injecting the pulse template into randomly triggered noise traces. The dashed line shows the 9.2~eV threshold set on the trigger energy estimator.}
    \label{fig:trig_efficiency}
\end{figure}  
Figure~\ref{fig:LowLambda} shows the energy distribution of a laser calibration dataset in which the average number of photons $\lambda$ absorbed in the Si substrate is of order 1 ($\lambda \sim 1$). The statistics of the dataset shown is large enough to extend the calibration to the fourth electron-hole pair peak, corresponding to a maximum energy of $\sim$400~eV.

The fill-in between the laser peaks can be explained via both charge trapping and impact ionization as charges propagate across the crystal~\cite{Ponce:2020}. In the former case, a charge is trapped in the crystal lattice, reducing the amount of phonon energy produced by shortening the drift length through the crystal. In the latter case, a charge kicks off a second loosely-bound unpaired charge increasing the total amount of energy collected. The charge trapping and impact ionization probabilities were evaluated for this detector by fitting the laser data with the model described in Ref.~\cite{Ponce:2020}. From these fits, we obtain a charge trapping of 12.7\% and an impact ionization of 0.6\% with the data acquired at TUNL. 

We employed the OF estimator described in Section~\ref{sec:expsetup} to evaluate the detector performance in the linear regime of the detector. We obtained a phonon energy resolution at the first electron-hole pair peak (corresponding to 101.95~eV for a NTL bias of 100~V) of 3.25(4)~eV, which corresponds to a charge resolution at the level of $\sim3$\% at 100~V bias. The measured phonon energy resolution was observed to be independent with respect to the applied NTL voltage below the point at which charge leakage begins to increase exponentially, as discussed in Refs.~\cite{Romani:2018,Amaral:2020}.

The baseline energy resolution was evaluated from a set of pseudo-random triggers on 0~V data. 
The amplitude was evaluated at the random trigger position with an optimum filter-based estimator without allowing the algorithm to search for the maximum.
A Gaussian fit results in a reconstructed energy resolution of $\sigma=2.65(2)$~eV, which is very closed to the value predicted in Sec.~\ref{sec:det_optimization}. 
We see a discrepancy between the baseline resolution and the resolution at the first electron-hole-pair peak, implying an additional source of energy smearing in the latter. This is likely due to surface absorption in the QETs~\cite{Hong:2020}.

The QET direct absorption is a known effect, and both the offset and variance of the laser peaks have been shown to correlate with the laser intensity~\cite{Hong:2020}. For 100~V laser data with $\lambda \sim 1$, the expected energy shift is of the order of 0.9~eV, which corresponds to less than a 1\% effect on the position of the first electron-hole-pair peak. This effect is taken into account during the calibration using laser data and thus will not impact reconstruction of events caused by a single bulk energy deposition.

The trigger efficiency was studied by injecting pulses into randomly triggered noise traces. The OF pulse template, which is the averaged laser pulse, was used as the shape of the injected pulses. A trigger time cut around the expected position of the injected pulse ($\pm3\sigma$ of the timing resolution which is equal to $\sigma=440$~ns for 15~eV events) ensures the correspondence between the injected pulse and the triggered one. The efficiency was calculated as the fraction of injected pulses that were triggered by the OF trigger algorithm. We achieved a threshold of 9.2~eV, which corresponds to 3.5$\sigma$ of the baseline resolution, while maintaining the trigger rate as low as 20~Hz. The resulting efficiency curve is shown in Fig.~\ref{fig:trig_efficiency}.

\begin{figure}[t]
    \centering
    \includegraphics[width=\columnwidth]{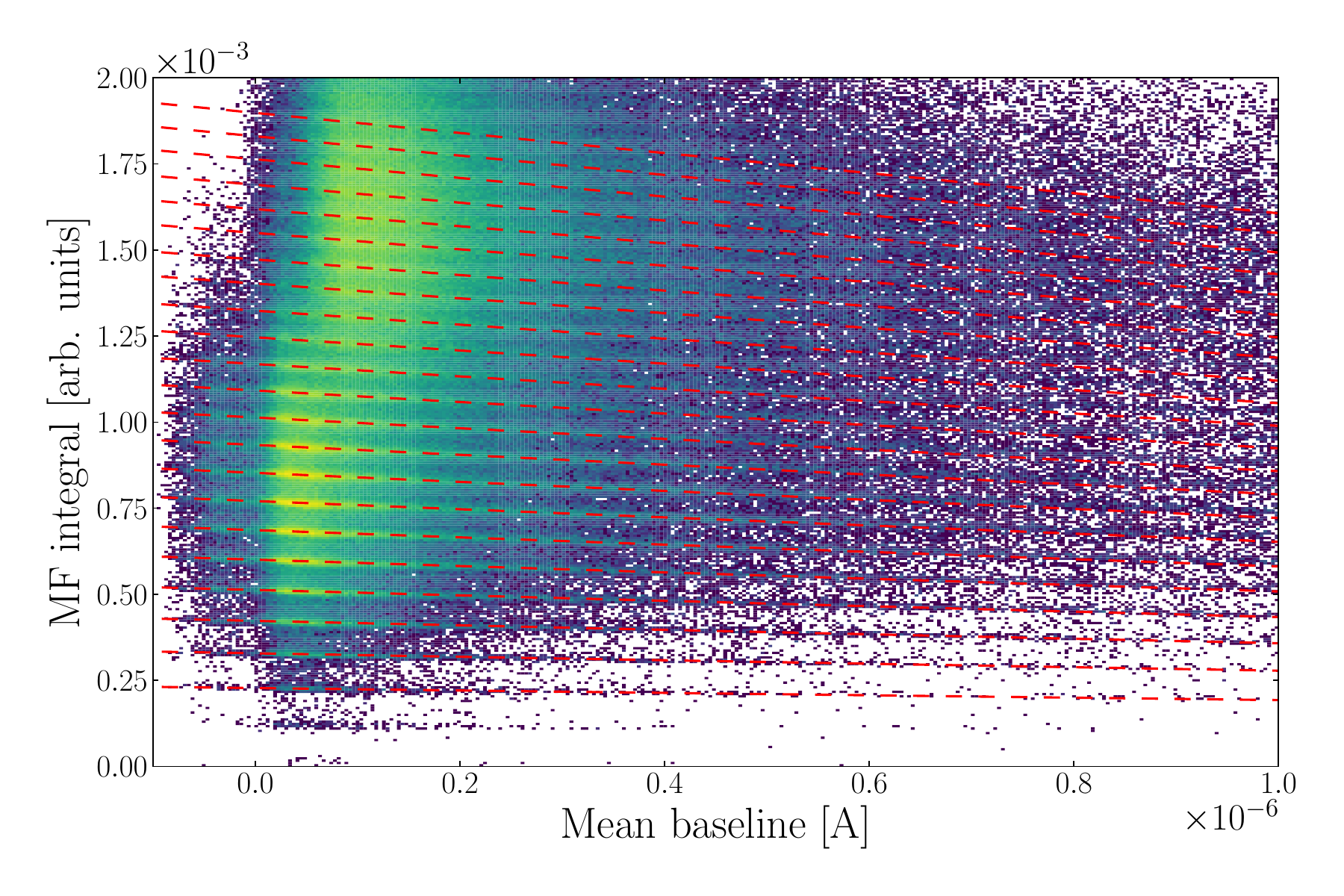}
    \includegraphics[width=\columnwidth]{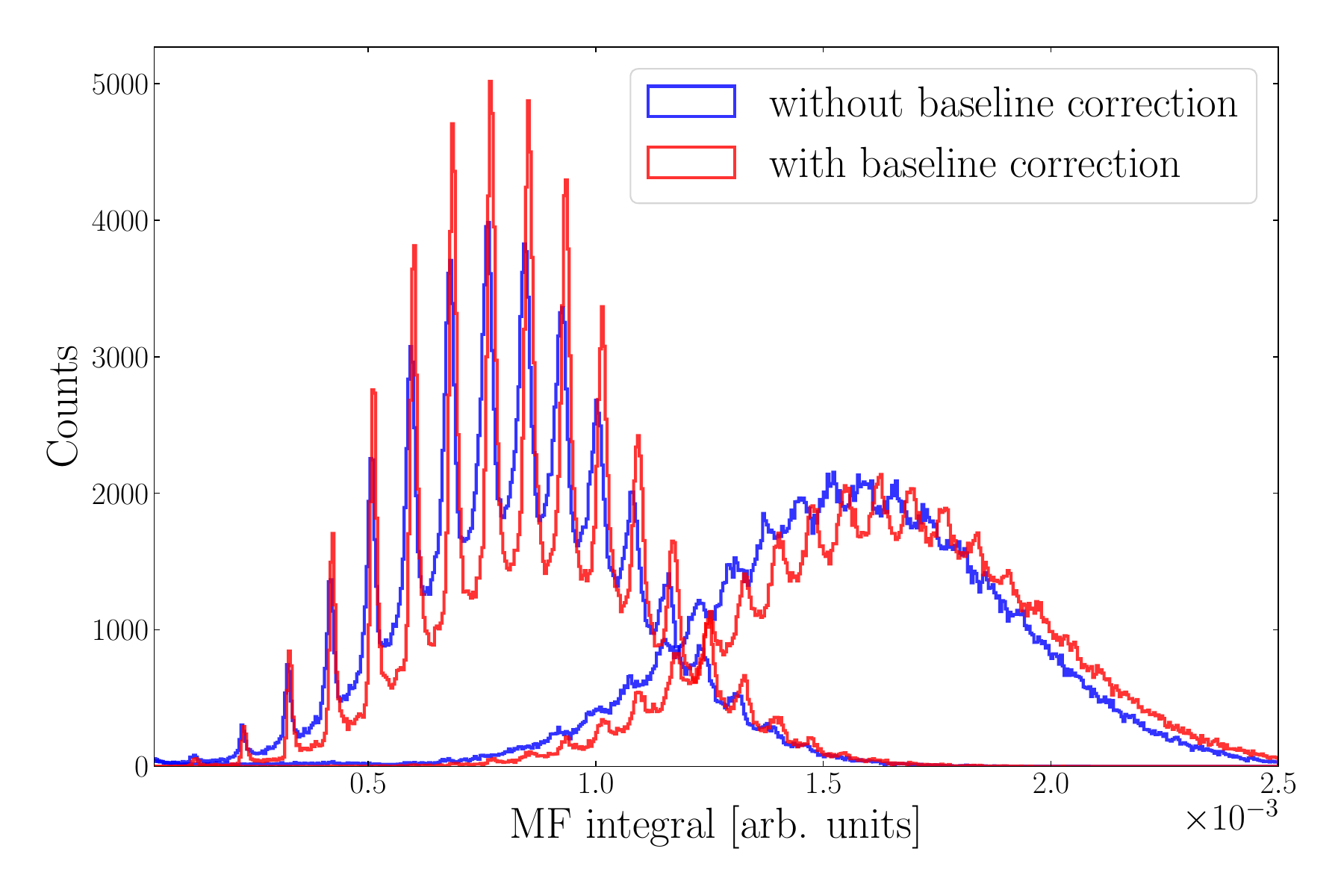}
    \caption{Top panel: 2D histogram of the matched-filter-based energy estimator as a function of the mean baseline for the laser data. Two data series with a different average number of photons were used, which is visible by the two clusters at low mean-baseline values. The red lines highlight the trend of the detector energy as a function of the mean baseline. Bottom panel: Laser spectrum before and after the mean-baseline gain correction for the two laser data series used in the correction.}
    \label{fig:meanbaseline}
\end{figure} 

\subsection{High-Energy Calibration}

Calibration of energies above the nominal linear region of $\sim$1~keV was accomplished by combining: (1) high-intensity laser data, up to 6~keV at HV bias; (2) an $^{55}$Fe source, which extends the calibration up to 120~keV by applying a voltage bias of 70~V; (3) data taken with a $^{57}$Co source without NTL bias. Past work demonstrated that laser data can be used to calibrate energies below 700~eV~\cite{Romani:2018,Agnese:2018,Amaral:2020}, as shown in the previous section, and other groups have used the $\sim$6~keV double peak from $^{55}$Fe, and associated 1.5~keV Al fluorescence, to calibrate the detector energy scale above 1~keV~\cite{Fink:2020b}. Here we demonstrate, for the first time, a combined approach to linearize the energy scale across four orders of magnitude in energy, combining the low-amplitude linear response region with the high-amplitude saturation region of the QET channels.

The first step was to model the response of the detector to the laser calibration signal at higher average photon number. The number of photons emitted by the laser was Poisson-distributed and was controlled by increasing the laser excitation current. The number of events populating the peaks (which are then used for the calibration) were reduced, due to the charge trapping and the impact ionization effects mentioned in the previous section. A longer acquisition time ($\sim 5$~hours) and a high laser rate ($\sim$101~Hz) were used to collect sufficient statistics for this first calibration step.

The high pulse rate, combined with a non-shielded cryostat operated in an above-ground facility, greatly increased the probability of pileup pulses. This caused the working point of the detector to shift, leading to a reduction in pulse height for a given energy deposit. The mean pre-pulse baseline, defined as the average value of 900-samples in the pre-pulse trace, directly measured the detector bias current, and was used to correct for this gain variation~\cite{Alessandrello:1998}. 
 
Figure~\ref{fig:meanbaseline} (top) shows the reconstructed pulse amplitude as a function of mean baseline for two data sets of laser data, demonstrating that the MF integral of each peak decreases as the mean baseline value increases. The correlation between laser peak positions in the mean baseline and amplitude plane has been approximated with a linear function and is shown in red for each electron-hole-pair peak in Fig.~\ref{fig:meanbaseline}. The mean-baseline correction was achieved by rotating the red lines around the zero-point on the mean-baseline axis, corresponding to the nominal detector baseline level. 

We rejected events above 1~$\mu$A in mean-baseline, limiting ourselves to the linear regime of this dependence and neglecting small nonlinear effects that are appreciable only over a larger mean-baseline range beyond 1~$\mu$A. Figure~\ref{fig:meanbaseline} (bottom) shows the laser spectrum before and after this correction; the improvement in the energy resolution and peak definition is evident. These laser data, acquired with NTL bias of 100~V and 250~V, provided a calibration up to 6~keV by using the first 24 peaks.

\begin{figure}[t]
    \centering
    \includegraphics[width=\columnwidth]{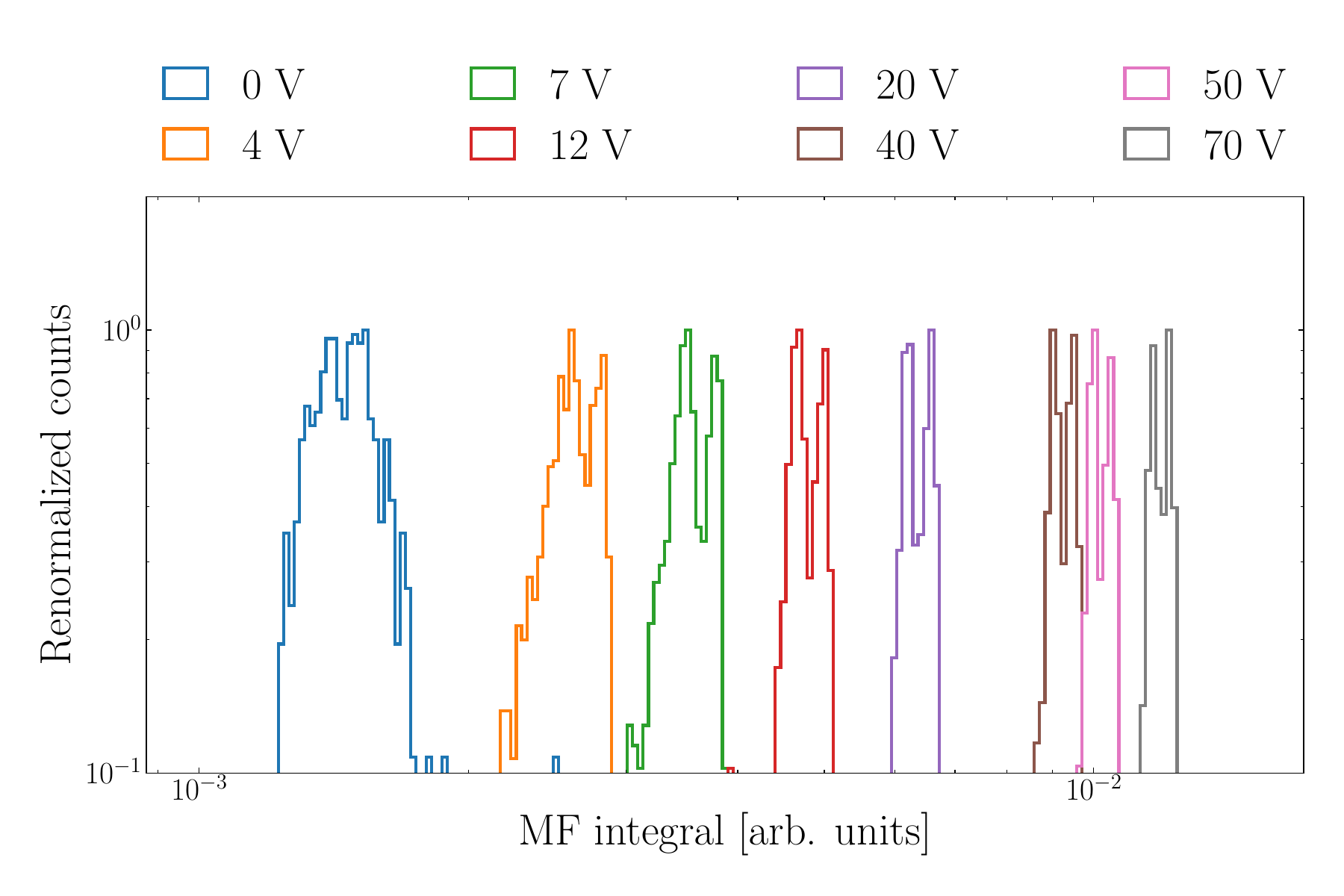}
    \includegraphics[width=\columnwidth]{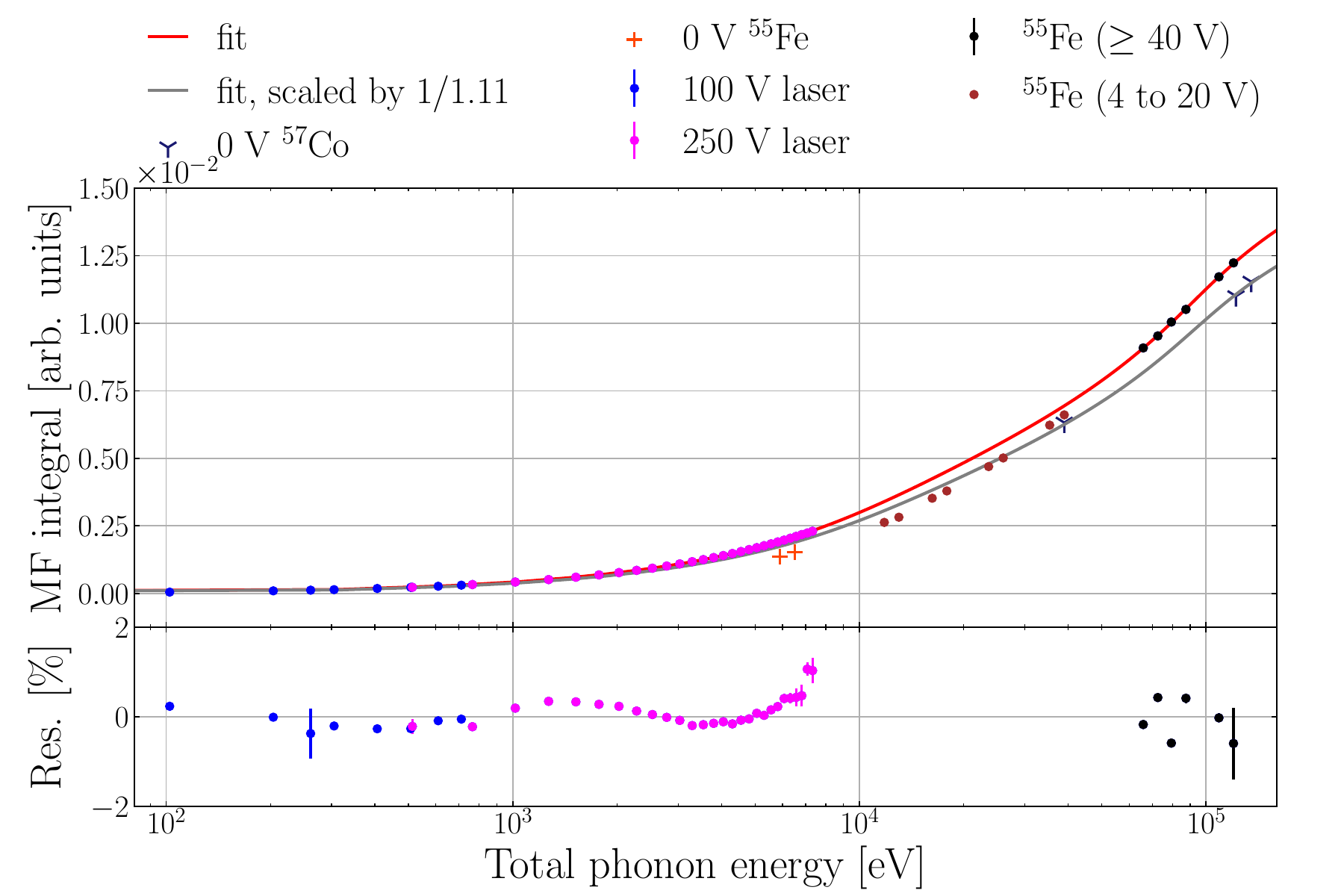}
    \caption{Top panel: $^{55}$Fe distribution for different NTL bias; the distributions are re-normalized by the maximum of the distribution. The two-peak structure corresponds to the $^{55}$Mn K$_{\alpha}$ and K$_{\beta}$ X-rays at 5.9 and 6.5~keV, respectively. Bottom panel: Combined calibration, including laser and $^{55}$Fe data.
    }
    \label{fig:FeAndcalib}
\end{figure}

The calibration at high energy used an external $^{55}$Fe source, which emits two $^{55}$Mn X-rays at 5.9 and 6.5~keV. The data were acquired at eight different NTL biases in order to uniformly cover the energy region between 6 and 120~keV. Figure~\ref{fig:FeAndcalib} (top) shows the measured $^{55}$Fe energy distributions used for this calibration. The use of a source outside of the cryostat produced an unusual event ratio between the K$_{\alpha}$ and K$_{\beta}$ lines caused by a decrease in attenuation of the X-rays with increasing energy.

We also incorporated the trapping and impact ionization effect to model the expected energy distribution of these peaks at high voltage. In the many charge limit, charge trapping and impact ionization effects can be included in the energy calibration using the relation:
\begin{equation}
    E_{ph}=E_{r}\cdot G_{\mathrm{NTL}} (1-0.5\cdot P_{\mathrm{CT}} +0.5\cdot P_{\mathrm{II}}), 
\end{equation}
where $P_{\mathrm{II}}$ and $P_{\mathrm{CT}}$ are the impact-ionization and charge-trapping probabilities, $G_{\mathrm{NTL}}=1+e\cdot V_{\mathrm{NTL}/\epsilon_{\gamma}}$ is the NTL gain and $E_{ph}$ and $E_{r}$ are the final phonon energy and the initial recoil energy. The factor 0.5 assumes that the charge trapping and impact ionization occur evenly across the detector. We expected a decrease in the energy scale of the order of 4.5\% by using the probabilities measured by the fit, as discussed in the previous section. This factor was included in the final energy calibration in Fig.~\ref{fig:FeAndcalib} (bottom). It was relevant for both calibration and background data at high voltage. This correction to the energy scale assumes that the trapping and impact ionization at the detector surface are the same as in the volume.

One finding from these data post-calibration was a mismatch between the calibration obtained with the laser source at high voltage and the calibration obtained with the $^{55}$Fe source at low voltage. The most likely mechanisms which could account for this discrepancy are: (1)~the NTL phonons have a different response with respect to the phonons generated by charge recombination; (2)~the penetration length of X-rays in Si ($\sim 30~\mathrm{\mu}$m) is not sufficient to reach the bulk and there is some signal degradation due to surface effects; (3)~the deposition of a single X-ray could generate local saturation in the sensor, because the $^{55}$Fe source was directed at the QET-instrumented face. The temperature distribution of the individual QETs can be strongly nonuniform due to a near-surface energy deposition causing only those QETs in the local vicinity of the deposition to saturate.

In the high-electric-field regime, (1) the charges are quickly drifted to the detector bulk, and (2) the phonon signal is dominated by NTL amplification such that the original energy deposition is negligible in comparison. This second point ensures that the phonon distributions with laser and $^{55}$Fe events are produced by the same mechanism. For these reasons, we only included the $^{55}$Fe calibration for nonzero voltage bias when the NTL effect accounts for more than 90\% of the expected phonon energy, corresponding to voltage biases in the range $40-70$~V. In this limit, a smooth energy reconstruction was possible for data acquired in the presence of strong NTL amplification.

We used an external $^{\mathrm{57}}$Co source with its two gamma rays at 122~keV and 136~keV and the 39-keV Compton edge in order to calibrate the data acquired without the NTL amplification.
The 0~V calibration data were fit using the same curve shape as for the HV data---a sixth order polynomial---multiplied by a scaling factor. The scaling factor was extracted from the fit and corresponds to 1/1.11. This curve is represented in Fig.~\ref{fig:FeAndcalib} in gray. The $^{55}$Fe data acquired at 0~V were not compatible with this curve, we suspect that this is due to the aforementioned local saturation and surface effects.

\begin{figure}[t]
    \centering
    \includegraphics[width=\columnwidth]{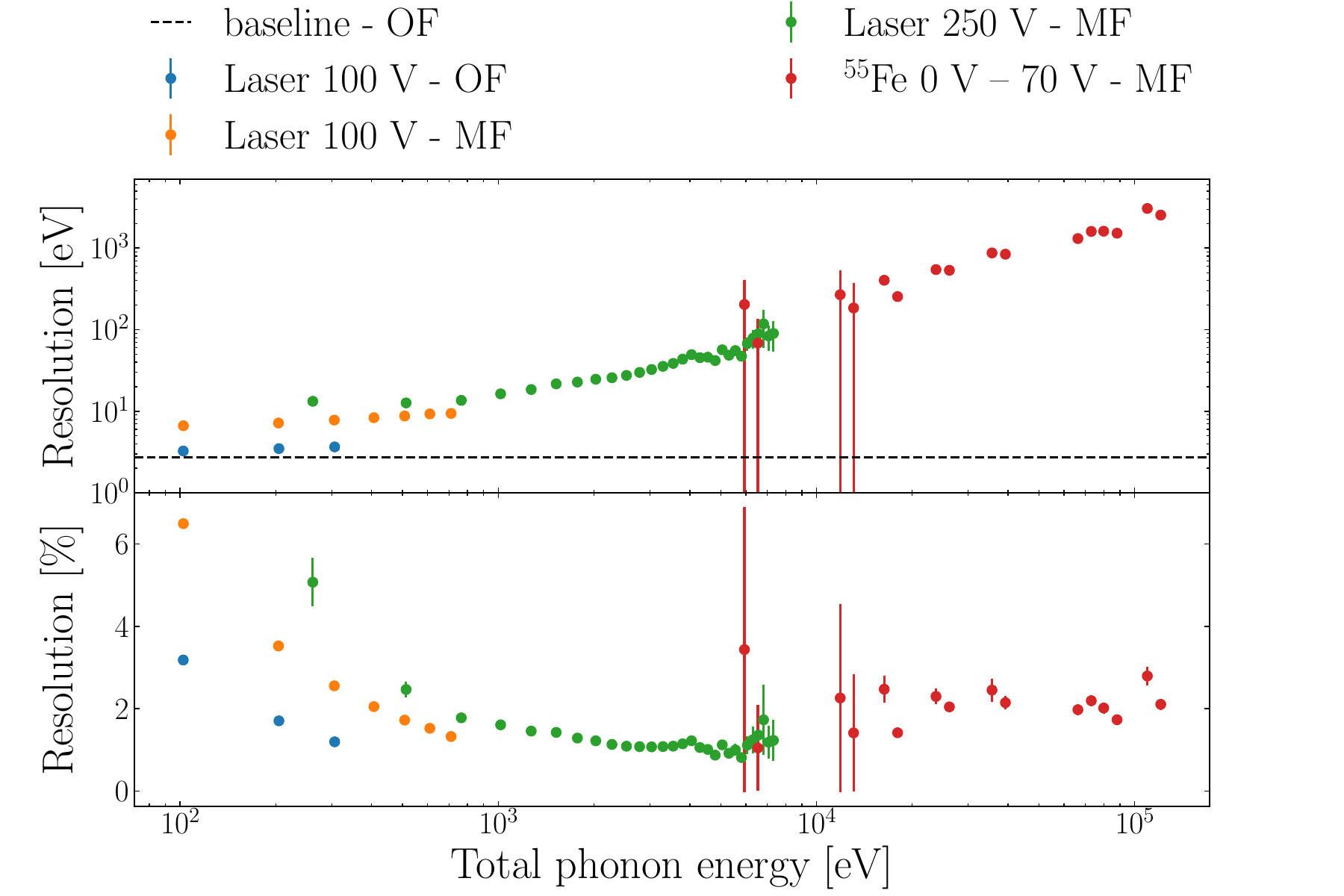}
    \caption{Energy resolution expressed as a function of the energy both for the OF and the MF energy estimators. The OF processing achieved the best energy resolution of $\sigma=3.25(4)$~eV for low energies. The MF estimator allowed us to obtain a fractional energy resolution less than 5\% for energies above 1~keV.}
    \label{fig:resolution}
\end{figure}

Figure~\ref{fig:resolution} shows the variation of the energy resolution as a function of the energy of the event. The OF energy estimator demonstrated an energy resolution of 3.25(4)~eV at 101.95~eV, as discussed in the previous section. The MF integral trades energy resolution for dynamic range, allowing us to probe much higher energies while maintaining an energy resolution lower than 5\%.


\section{Conclusion}\label{sec:conclusion}
We have presented a second-generation single-charge-sensitive detector. 
The detector design was optimized to improve the energy resolution and to enlarge the dynamic range of the detector described in Refs.~\cite{Romani:2018, Agnese:2018}. This detector also achieved a total energy collection efficiency in excess of 29\%, the highest yet measured for a phonon calorimeter. Further characterization of our readout circuit is needed to more precisely measure energy efficiency. This detector demonstrated a baseline resolution of 2.65(2)~eV, which allowed us to set a threshold of 9.2~eV while accepting a $\sim$20-Hz rate due to background and noise events. The detector calibration was extended up to 120~keV thanks to the use of an energy estimator based on the pulse area. The extension of the energy calibration into tens of keV was essential for the ionization yield measurement carried out at TUNL. 

Continued studies of the detector discussed in this paper are underway at NEXUS. In particular, the complex impedance studies used for the noise model were limited by systematics on the TES shunt resistance and poor impedance matching of the test signal, and further studies will allow us to refine these measurements. These systematics prevent us from making quantitative statements about the monolithic TES thermal model used in the noise analysis, despite growing evidence that a more complex thermal model is needed for low-$T_c$ QETs~\cite{Fink:2020}. Noise and complex impedance as a function of bath temperature and bias point will help us better understand the internal thermal degrees of freedom of the QETs.

Finally, Fig.~\ref{fig:detmodel} suggests that lower resolution is achievable with the same QET design by simply reducing TES length, at the expense of overall dynamic range. Multiple designs were fabricated closer to this resolution optimum in the same batch as this detector and are also under test as of this writing. Comparison of the efficiency and NEP in these designs will allow us to better quantify the impact of phonon losses on the energy efficiency of these devices. If phonon losses are minimal, we expect multiple of these designs to approach 1~eV baseline resolution. Fabrication of lower $T_c$ devices like those in Ref~\cite{Fink:2020} will allow these devices to achieve sub-eV resolution and single optical photon resolution at 0~V.

\begin{acknowledgments}
We would like to thank Martin Huber for providing the SQUIDs used in these measurements, Suhas Ganjam for help in initial detector mask design, and Jillian Gomez for help with initial commissioning of the NEXUS cryostat. We would also like to thank Paul Brink, Blas Cabrera, Sunil Golwala, Belina von Krosigk, and Betty Young for feedback on the draft their support of this work. We gratefully acknowledge support from the  U.S.  Department  of  Energy  (DOE)  Office  of  High Energy  Physics  and  from  the  National  Science Foundation  (NSF). This work was supported in  part  under NSF Grants No. PHY-1809730, as well as by the Deutsche Forschungsgemeinschaft (DFG) under Project No. 420484612 and Germany’s Excellence Strategy - EXC 2121 ``Quantum Universe" – 390833306. Parts of this document were prepared using the resources of the Fermi National Accelerator Laboratory (Fermilab), a U.S. Department of Energy, Office of Science, HEP User Facility. Fermilab is managed by Fermi Research Alliance, LLC (FRA), acting under Contract No.~DE-AC02-07CH11359. Pacific Northwest National Laboratory (PNNL) is operated by Battelle Memorial Institute for the U.S. Department of Energy under Contract No.~DE-AC05-76RL01830.
\end{acknowledgments}

\appendix
\section{QET Dynamic Range}\label{app:DR}
To determine the dynamic range of a TES-based sensor, we want to calculate the ratio of saturation energy (energy required to drive the TES normal) to energy resolution. Given that the observable is TES current, we can calculate the ratio of the pulse height for an impulse of total energy equal to the energy resolution to the maximum current change from the bias point, which yields a dimensionless ratio useful for calculating current or energy quantities.

Consider an ideal voltage-biased TES, assuming the operating point is much greater than the shunt resistance in the TES bias loop~\cite{Irwin:2005}. We find that the saturation current scales as
\begin{equation}
    I_{sat} \approx V_{b}\left[\frac{1}{R_0}-\frac{1}{R_n}\right] = \frac{V_b}{R_n}\left[a^{-1}-1\right]
\end{equation}
where the bias resistance $R_0=aR_n$, $R_n$ is the normal state resistance, and $V_b$ is the TES bias voltage. The equilibrium bias condition tells us that Joule power and thermal conductance power from the TES to the crystal substrate will balance, which allows us to calculate bias voltage as
\begin{equation}
    V_b = \sqrt{aR_n\Sigma \frac{v_\tes}{\zeta_\tes}T_c^n}
\end{equation}
giving us an equation for saturation current:
\begin{equation}
    I_{sat}  \approx \sqrt{\frac{a}{R_n}\Sigma \frac{v_\tes}{\zeta_\tes} T_c^n}\left[a^{-1}-1\right]
\end{equation}
This clearly has similar scalings to the resolution.

In the small signal limit, we want to calculate the current amplitude for an injection pulse of energy equal to the energy resolution. We assume that the phonon pulse follows the simple exponential form
\begin{equation}
    P(t) = \frac{\epsilon\sigma}{\tau_{ph}}e^{-t/\tau_{ph}}
\end{equation}
and the Green's function response of the TES has the form~\cite{Irwin:2005}
\begin{equation}
\delta I(t)=\frac{\mathcal{L}}{1+\beta}\frac{\Delta E}{V_b\tau}e^{-t/\tau_-}
\end{equation}
where we have implicitly assumed that the rise time is much shorter than the fall-time of the TES (we assume we are operating in the limit of low inductance). If we write $\Delta E = P(t'-t)\Delta t'$, we can derive the QET response function by convolving the two pulses
\begin{equation}
\begin{split}
\delta I(t)=&\frac{\mathcal{L}}{1+\beta}\frac{1}{V_b\tau}\epsilon\sigma \\
& \frac{1}{1-\tau_{ph}/\tau_{-}}\left[e^{-t/\tau_{-}}-e^{-t/\tau_{ph}}\right] 
\end{split}
\end{equation}

This function has two limits: (1) when the TES response time is much larger than the phonon response time, the right-most term reduces to the TES Green's function, which has an amplitude given by the coefficient; (2) in the limit that the phonon response time is much larger than the TES response time, the amplitude is corrected by the fall time ratio. Solving precisely for maximum amplitude of the time-dependent part of this function, we find the formula for maximum amplitude
\begin{equation}
    I_{\sigma} = \epsilon\sigma\frac{\mathcal{L}}{1+\beta}\frac{1}{V_b\tau}\left(\frac{\tau_{ph}}{\tau_{-}}\right)^{\frac{-\tau_{ph}}{\tau_{-}-\tau_{ph}}}
\end{equation}
where we can see that a long phonon fall-time reduces the maximum pulse height for the same TES response.

Finally, we can calcuate dynamic range by taking the ratio of saturation current to pulse amplitude
\begin{align}
    & DR = \frac{I_{sat}}{I_{\sigma}} \\
      &= \frac{1}{\epsilon\sigma}\frac{V_b^2}{R_n}\left[a^{-1}-1\right]\frac{1+\beta}{\mathcal{L}}\tau \left(\frac{\tau_{ph}}{\tau_{-}}\right)^{\frac{\tau_{ph}}{\tau_{-}-\tau_{ph}}} \\
      &= \frac{1}{\epsilon\sigma}\frac{\Sigma v_\tes T_c^n}{\zeta_\tes}\left[1-a\right]\frac{1+\beta}{\mathcal{L}}\tau \left(\frac{\tau_{ph}}{\tau_{-}}\right)^{\frac{\tau_{ph}}{\tau_{-}-\tau_{ph}}} \\
      &= \frac{1}{\epsilon\sigma}\frac{f_{sc}c_{W}v_\tes T_c^2}{n\zeta_\tes}\left[1-a\right]\frac{1+\beta}{\mathcal{L}} \left(\frac{\tau_{ph}}{\tau_{-}}\right)^{\frac{\tau_{ph}}{\tau_{-}-\tau_{ph}}}\label{eq:DR}
\end{align}
where this last step follows from the $T_c$ dependence of $\tau$~\cite{Irwin:2005,Kurinsky:2018},
\begin{equation}
    \tau = \frac{f_{sc}c_W}{n\Sigma}T_c^{2-n}.
\end{equation}
Here, $f_{sc}$ is the superconductivity enhancement to the specific heat and $c_W$ is the normal state specific heat.

This last scaling, if we assume the TES transition shape is invariant with $T_c$ and TES volume, shows us how to maximize dynamic range of a device without degrading resolution. If we fix device $T_c$ and hold resolution constant by definition, we find that from Eq.~\ref{eq:t3}, $\sigma \propto \frac{\sqrt{v_\tes}}{\epsilon}$; so for a fixed resolution, $DR \propto \sqrt{v_\tes}$. Scaling up total TES volume will only improve dynamic range, without degrading resolution, if we can also increase device efficiency as the square root of volume enhancement.

While Eq.~\ref{eq:DR} is exact in the case that $T_c$ and geometry dependence of the various device parameters are known, these scalings only hold in the specific limit that we can be reasonably certain that TES response ($\mathcal{L}$ and $\beta$) will not change with efficiency and volume scaling. In this paper, design changes were largely limited to the size and number of QETs and thus we could reasonably model efficiency and volume as independent of TES response, benchmarking TES constants to previous devices such as the QP.4 detector discussion in the text. We should note, however, that the fully general calculation should add back in considerations for TES rise time and more complex phonon response characteristics, and thus this scaling serves as more of a general design guide than a precise calculation.

\section{QET Efficiency Modeling}\label{app:eff}

A complete description of the energy efficiency model can be found in Section 3.4 of Ref.~\cite{Kurinsky:2018}. Here, we briefly summarize the key features of this model and discuss how further refining the measured efficiency of this device, and comparable designs, can inform this model.

The total energy efficiency for converting phonon energy into the TES (that is subsequently detected) can be split into four main components, as illustrated in the top panel of Fig.~\ref{fig:detmodel}:
\begin{enumerate}
    \item Phonon collection efficiency $\epsilon_{ph}$, the probability that an initial phonon is absorbed by an Al fin;
    \item Phonon to quasiparticle conversion efficiency $\epsilon_{qp}$ for a phonon absorbed in the fin;
    \item QP collection efficiency $\epsilon_{coll}$ for QPs concentrated into the trapping regions;
    \item Trapped QP to TES thermal energy conversion efficiency, $\epsilon_{trap}$.
\end{enumerate}
All efficiencies are applied on a per-phonon or per-quasiparticle basis and are assumed to be energy independent. The total efficiency used in the resolution calculation is thus $\epsilon = \epsilon_{ph}\epsilon_{qp}\epsilon_{coll}\epsilon_{trap}$. For a practical device, only the first and third efficiencies are readily tunable through design optimization; the conversion efficiencies are largely material-defined rather than geometric.

Overall phonon collection inefficiency can be further split into phonon losses in the bulk, at surfaces, and to non-instrumented absorption. For sufficiently pure crystals, bulk losses are negligible and surface effects dominate, as discussed in Ref.~\cite{Griffin:2020}. For the current device, phonon losses are further minimized with high surface coverage. This enhances the probability for phonon absorption before their energy drops below the Al band gap by down conversion on crystal surfaces. In addition, there is very little uninstrumented absorbing surface, as the bias rails of this device are integrated into the QET fins. Thus the dominant phonon losses are expected to be in the back-side grid and at the detector side-walls. These losses are further mitigated by (1) making the backside grid only 30~nm thick, compared to the 600~nm thick absorbing fins, allowing phonons to be reflected back into the substrate before breaking Cooper pairs, and (2) using a large aspect ratio device to minimize total side-wall area. Our model suggests that these design choices are consistent with $\epsilon_{ph}\gtrsim 95$\%, given that phonon losses are assumed to be fairly negligible in this limit (assuming that there are no bulk or sidewall phonon losses) and that the fraction of phonons absorbed in the backside grid scales linearly with thickness\footnote{We can model the backside absorption fraction, for equal surface coverage, as $f_{loss}=f_{front}/(f_{front}+f_{back})\sim$5\% for the thicknesses used in this design. When the backside grid has a lower coverage than the QET pattern, as is the case with this detector, this should be an upper bound on total phonon loss.}.

The largest fixed efficiency reduction comes from the limited efficiency of phonon to quasiparticle conversion in the Al fins, referred to as Kaplan down-conversion. Detailed studies of the energy dependence of this process can be found in e.g. Refs.~\cite{Kaplan:1976,Guruswamy:2015}. For typical phonon energies many times larger than the superconducting gap energy, this process is limited to an efficiency of around $50-60$\%. Close to the gap, the efficiency increases due to the reduced fraction of the energy which can be released as phonons. This means that a phonon sensor using a superconducting absorber is fundamentally limited by the mismatch between the phonon energy distribution and superconducting gap. We take $\epsilon_{qp}\sim50$\% as an upper limit on the efficiency of our sensors.

A related efficiency is the down-conversion of QPs to phonons plus normal electrons in the TES. In principle, this efficiency can be as high as 100\% if the phonon energy can be contained to the TES, but there are losses both in the transport regions between the trap and TES, and during the phonon emission process, to the TES. Experiments measuring efficiency difference between events absorbed in the fin, and directly by the TES, imply that this efficiency is roughly 62\% for simple trap designs~\cite{Kurinsky:2018}. A dedicated study of the trap design used by more recent QETs has yet to be fully characterized, but is expected to be higher. If we take this efficiency as a bound, this implies that, for perfect QP and phonon collection, our devices can at best expect an efficiency of $30-40$\%, limited by the energy conversion efficiency of the phonon to QP to phonon process.

The final consideration, which can be highly optimized, is the QP collection. Past studies have shown that QPs in high-quality Al fins have diffusion lengths on the order of hundreds of microns, but that the collection fraction of QPs is a function of the Al fin length, fin thickness, and trap geometry~\cite{Yen:2014,Yen:2016}. This can be understood as a quasi two-dimensional diffusion problem, in which the diffusion length is also a function of the film thickness (in the limit that the fins are much longer than this thickness), and the collection at the interface depends on the transmission probability through that interface as well as the probability of diffusing to the interface. A detailed discussion can be found in Ref.~\cite{Kurinsky:2018}; for fins shorter than 100~$\mu$m, all geometries can expect collection efficiency greater than 75\%. As the TES gets longer, the Al area can be split into more QET fins with individual Al/W trap areas. The diffusion in the fins then becomes more 1-dimensional, which increases the effective diffusion length and leads to more efficient QP collection for a fixed fin length.

Bounding the QP collection efficiency thus implies that, for designs with quasi-1D QP diffusion and short fin lengths, we will expect efficiencies on the order of $20-30$\%. The quoted efficiency in this paper, a lower bound of 29\%, suggests that these assumptions are realistic. A more precise measurement of the efficiency of multiple detectors with the same QET design, but different surface area scalings, will help better quantify the remaining uncertainty in the model.

\bibliographystyle{revtex-4-1.bst} 
\bibliography{refs} 
\end{document}